\documentclass[10pt]{wlscirep}

\usepackage[utf8]{inputenc}
\usepackage[T1]{fontenc}
\usepackage{siunitx}
\usepackage{chemist}

\usepackage{makecell}
\usepackage{csquotes}

\DeclareSymbolFont{mymathvariables}{T1}{lmr}{m}{it}
\SetSymbolFont{mymathvariables}{normal}{T1}{lmr}{m}{it}
\DeclareMathSymbol{\varv}{\mathalpha}{mymathvariables}{`v}

\title{3D characterisation of individual grains of coexisting high-pressure \chemform{H_2O} ice phases by time-domain Brillouin scattering}

\author[1,+]{Sathyan Sandeep}
\author[1]{Théo Thréard}
\author[1,+]{Elton De Lima Savi}
\author[1]{Nikolay Chigarev}
\author[2]{Alain Bulou}
\author[1]{Vincent Tournat}
\author[3]{Andreas Zerr}
\author[1,*]{Vitalyi E. Gusev}
\author[1,4,*]{Samuel Raetz}

\affil[1]{Laboratoire d'Acoustique de l'Université du Mans (LAUM), UMR 6613, Institut d'Acoustique - Graduate School (IA-GS), CNRS, Le Mans Université, France}
\affil[2]{Institut des Molécules et Matériaux du Mans (IMMM), UMR 6283, CNRS, Le Mans Université, France}
\affil[3]{Laboratoire de Sciences des Procédés et des Matériaux (LSPM-CNRS UPR-3407), Université Sorbonne Paris Nord (USPN), 93430 Villetaneuse, France}
\affil[4]{Associate member of Laboratoire Cogitamus, Av. O. Messiaen, 72085 Le Mans cedex 9, France}

\affil[*]{vitali.goussev@univ-lemans.fr, samuel.raetz@univ-lemans.fr}
\affil[+]{these authors contributed equally to this work}


\begin{abstract}
Time-domain Brillouin scattering uses ultrashort laser pulses to generate coherent acoustic pulses of picoseconds duration in a solid sample and to follow their propagation in order to image material inhomogeneities with sub-optical depth resolution. The width of the acoustic pulse limits the spatial resolution of the technique along the direction of the pulse propagation to less than several tens of nanometres. Thus, the time-domain Brillouin scattering outperforms axial resolution of the classical frequency-domain Brillouin scattering microscopy, which uses continuous lasers and thermal phonons and which spatial resolution is controlled by light focusing. The technique benefits from the application of the coherent acoustic phonons, and its application has exciting perspectives for the nanoscale imaging in biomedical and material sciences. In this study, we report on the application of the time-domain Brillouin scattering to the 3D imaging of a polycrystal of water ice containing two high-pressure phases. The imaging, accomplished via a simultaneous detection of quasi-longitudinal and quasi-shear waves, provided the opportunity to identify the phase for individual grains and evaluate their crystallographic orientation. Monitoring the propagation of the acoustic waves in two neighbouring grains simultaneously provided an additional mean for the localisation of the grain boundaries.
\end{abstract}
\begin{document}

\flushbottom
\maketitle

\thispagestyle{empty}

\section*{Introduction}
The application of all-optical techniques for characterisation of materials at high pressures is due to the transparency of diamonds which are used for compression of materials in a diamond anvil cell (DAC)\cite{Bassett2009_v29_p163-186,Eremets1996_v_p,Jayaraman1983_v55_p65--108}. Typically, samples compressed in a DAC have shapes of thin disks of 50-400~\si{\micro\metre} in diameter and of 10-50~\si{\micro\metre} in thickness, laterally supported by metallic gaskets. Examples of the fruitful application in the high pressure research of the Raman scattering\cite{Jayaraman1983_v55_p65--108,Polian2003_v34_p633--637} and of the classical frequency-domain Brillouin scattering (FDBS)\cite{Polian2003_v34_p633--637,Polian1991_v286_p181--193}, employing the interaction of photons with thermal phonons, are well known\cite{Jayaraman1983_v55_p65--108,Polian2003_v34_p633--637, Polian1991_v286_p181--193}. Quite recently, another opto-acousto-optical technique based on the interaction of light with a propagating density perturbation (an acoustic pulse) caused by coherent non-thermal acoustic phonons, was introduced into high-pressure experiments\cite{Decremps2008_v100_p035502,Armstrong2008_v92_p101930}. The technique considered here had been earlier pioneered for materials characterisation at ambient conditions under the name of picosecond acoustic interferometry\cite{Thomsen1986_v60_p55--58,Grahn1989_v25_p2562--2569}, while currently the names of time-resolved or time-domain Brillouin scattering (TDBS) have become more common\cite{Gusev2018_v5_p031101}. In this technique, based on the principles of pump-probe time-resolved experiments, the pump laser pulses of the picoseconds-femtoseconds duration are used to launch in a compressed sample coherent acoustic pulses (CAPs) of picosecond duration via one of the possible mechanisms of optoacoustic conversion\cite{Akhmanov1992_v35_p153--191,Ruello2015_v56_p21--35}. The probe laser pulses of picoseconds-femtoseconds duration, launched with adjustable time delays after the pump pulses, are used to follow in time the propagation of the CAPs due to the probe light scattering by the CAPs via opto-acoustic (photoelastic) effect\cite{Fabelinskii1968_v_p,Dil1982_v45_p285--334}. As the Brillouin light scattering takes place only in the volume occupied by the CAP, the depth spatial resolution of the TDBS technique along the path of the CAP can be sub-optical because typical width of a 10 ps duration CAP is shorter than 50 nm\cite{Gusev2018_v5_p031101,Steigerwald2009_v94_p111910,Mechri2009_v95_p091907,Lomonosov2012_v6_p1410--1415}. Thus, the depth resolution of the TDBS technique can outperform that of the FDBS microscopy which, in the absence of a control over the thermal phonons spatial localisation, is limited by the light focusing\cite{Scarcelli2008_v2_p39--43,Scarcelli2015_v12_p1132--1134}. Up to now, the record spatial depth resolution in FDBS has been reported in the confocal optical geometry: $\sim 0.5 \times 0.5 \times 2$~\si{\micro\metre\cubed}, yet at atmospheric pressure only\cite{Scarcelli2008_v2_p39--43,Scarcelli2015_v12_p1132--1134}. Application of the confocal optical FDBS microscopy to samples compressed in a DAC is technically demanding, if ever possible. In common FDBS experiments on samples compressed in a DAC, the Brillouin scattering takes place in the volume, which axial (depth) dimension is controlled either by the sample thickness or by the light penetration depth. In both cases, the FDBS technique does not provide any depth-resolved information, \textit{i.e.}, any possible variation in the properties with depth, from one culet to the opposite. The application of the TDBS depth-profiling and imaging therefore provides opportunities and advantages earlier unavailable in the solid state research at high pressures.

The TDBS technique was applied for the depth-profiling of polycrystalline water ice and solid argon up to megabar pressures\cite{Nikitin2015_v5_p9352,Kuriakose2016_v69_p259--267}, revealing in these granular assemblages not only \si{\micro\metre}-sized textures along the sample surface but also, for the first time, sub-\si{\micro\metre}- and \si{\micro\metre}-sized inhomogeneities along the sample axis, which are currently inaccessible for the FDBS technique. The TDBS technique was then applied for the quantitative characterisation of elastic inhomogeneities in water ice\cite{Kuriakose2017_v96_p134122} and solid argon\cite{Raetz2019_v99_p224102}, with a demonstrated better precision than by FDBS. The principle of this advantageous application of the TDBS technique in comparison with the FDBS one relies on the much better axial resolution of the first one. Currently, if single crystals are not accessible, which is the case for a big majority of high pressure (and/or high temperature) phases, elastic constants are extracted from the values of the maximum and minimum frequencies measured in the polycrystalline sample by the Brillouin scattering, under the assumption that they correspond to the theoretical values in the single crystals. Yet, these theoretical values are experimentally inaccessible unless the Brillouin scattering is conducted inside the individual crystallites. The number of crystallites in the Brillouin scattering volume is unfortunately commonly increasing with the increase of the tested volume. This means that the larger the scattering volume is, \textit{i.e.} the worse the spatial resolution of the Brillouin scattering technique is, the less is the chances that the measured Brillouin frequency closely approaches one of its extremal theoretical values. From the viewpoint of quantitative measurements, the TDBS therefore drastically outperforms the FDBS because of the significantly increased depth spatial resolution of the measurements permitting, in the particular cases of the rather thick crystallites, their individual examination in a DAC\cite{Kuriakose2017_v19_p053026}. In a more general case of several crystallites inside the Brillouin scattering volume, the combination of sub-\si{\micro\metre} axial resolution and lateral scanning in TDBS dramatically increases the number of measurements and the statistical probability to closely access the maximal and minimal longitudinal sound velocities, which then could be attributed, with a high degree of confidence, to the maximal and minimal possible sound velocities in single crystals. This permits to reliably evaluate single crystal elastic moduli $C_{ij}$ even though single crystals are technologically inaccessible\cite{Kuriakose2017_v96_p134122,Raetz2019_v99_p224102}. As a consequence, the results\cite{Kuriakose2017_v96_p134122} for water ice implied, for example, that the ionic phase ice X should first appear at pressures exceeding 80~GPa, while the transition from ice VII to ice X was earlier predicted to occur below 60~GPa, e.g. Refs. \citeonline{Polian1984_v52_p1312--1314,Loubeyre1999_v397_p503--506,Goncharov2006_v35_p899}. Another ability of the TDBS technique is monitoring of transient processes at high pressures such as imaging of the laser-induced spatial displacement of the boundary between two \chemform{H_2O} ice phases, ices VI and VII, on the time scale from hours to days\cite{Kuriakose2017_v19_p053026}.

Although the above-mentioned results\cite{Nikitin2015_v5_p9352,Kuriakose2016_v69_p259--267,Kuriakose2017_v96_p134122,Raetz2019_v99_p224102} have already demonstrated the efficiency of the 2D TDBS-based imaging for qualitative and quantitative characterisation of materials elasticity and phase transitions, it is not sufficient for visualisation and examination of texture and its evolution in time or upon compression. To do so, a 3D imaging of sample inhomogeneities upon compression is  required. The 3D TDBS imaging has been first realised at ambient conditions in animal and biological cells\cite{Danworaphong2015_v106_p163701,Perez-Cota2015_v54_p8388}, which are weakly inhomogeneous objects where regions having distinct elastic and/or optical parameters are separated by curved interfaces\cite{Danworaphong2015_v106_p163701,Perez-Cota2015_v54_p8388,Perez-Cota2016_v6_p39326}. It can be expected that in a compressed polycrystalline solid the neighbour regions with different properties or just different orientations, \textit{i.e.} crystallites, are separated by locally plane interfaces, \textit{i.e.} grain boundaries. For characterisation of evolution of the sample polycrystallinity upon compression, it is highly desirable to monitor it starting from the single crystal state. It is well known that it is possible to grow large single crystals in a DAC at low pressures, e.g. upon crystallisation of a liquid under slow compression\cite{Bassett2009_v29_p163-186}. It is even easier to prepare a polycrystal composed of the grains with spatial dimensions significantly exceeding the lateral resolution of the TDBS set-up. As already mentioned above, the lateral resolution in the TDBS imaging is commonly controlled by focusing of the probe laser pulses to \si{\micro\metre}- or sub-\si{\micro\metre}-radius spots. Accordingly, the 3D TDBS imaging can be realised for polycrystals containing initially grains with dimensions exceeding $\sim$10~\si{\micro\metre} and follow changes in the grain size and shape accompanied by transformations of the grain boundaries with increasing pressure until the characteristic grain dimension decreases below 1~\si{\micro\metre}, as reported in Ref.~\citeonline{Nikitin2015_v5_p9352}, for example, or even approaches the nanometres scale. At ambient conditions, TDBS experiments on polycrystalline materials with large grain sizes were already reported, and it was found that the TDBS imaging inside the individual grains is straightforward\cite{Lejman2014_v5_p4301,Lejman2016_v7_p12345,Khafizov2016_v112_p209--215,Wang2019_v166_p34--38,Wang2020_v11_p1597}. In particular, the simultaneous monitoring of the propagation of quasi-longitudinal acoustic (LA) and quasi-transverse acoustic (TA) pulses inside a grain by TDBS was demonstrated\cite{Lejman2014_v5_p4301,Lejman2016_v7_p12345}. Taking advantage of this feature, the TDBS technique provided opportunity to determine the orientation of an individual grain boundary first in 2D geometry\cite{Khafizov2016_v112_p209--215} and later in 3D geometry\cite{Wang2019_v166_p34--38}. Finally, the simultaneous TDBS imaging with LA and TA pulses revealed the crystallographic orientation of the individual grains inside a polycrystal\cite{Wang2019_v166_p34--38,Wang2020_v11_p1597}.

Here, we report on the extension of several concepts developed recently in the TDBS experiments at ambient conditions to the 3D TDBS imaging at high pressures in a DAC. In particular, we demonstrate the first experimental observation by TDBS of coherent TA pulses at high pressures. The TDBS imaging with both TA and LA pulses provided the opportunity to distinguish the water ice VII from the water ice VI at 2.15~\si{\giga\pascal} pressure, where they coexist at room temperature. The presence of TA pulses was also crucial in obtaining the orientation of the crystallographic axes of the individual ice grains relative to the DAC axis. In addition, we report on the detection of particular TDBS signals, corresponding to the propagation of LA pulses in two neighbour grains simultaneously, the observation of which provides an original opportunity to localise the boundaries between the grains.

\section*{Results}
\begin{figure}[ht]
\centering
\includegraphics{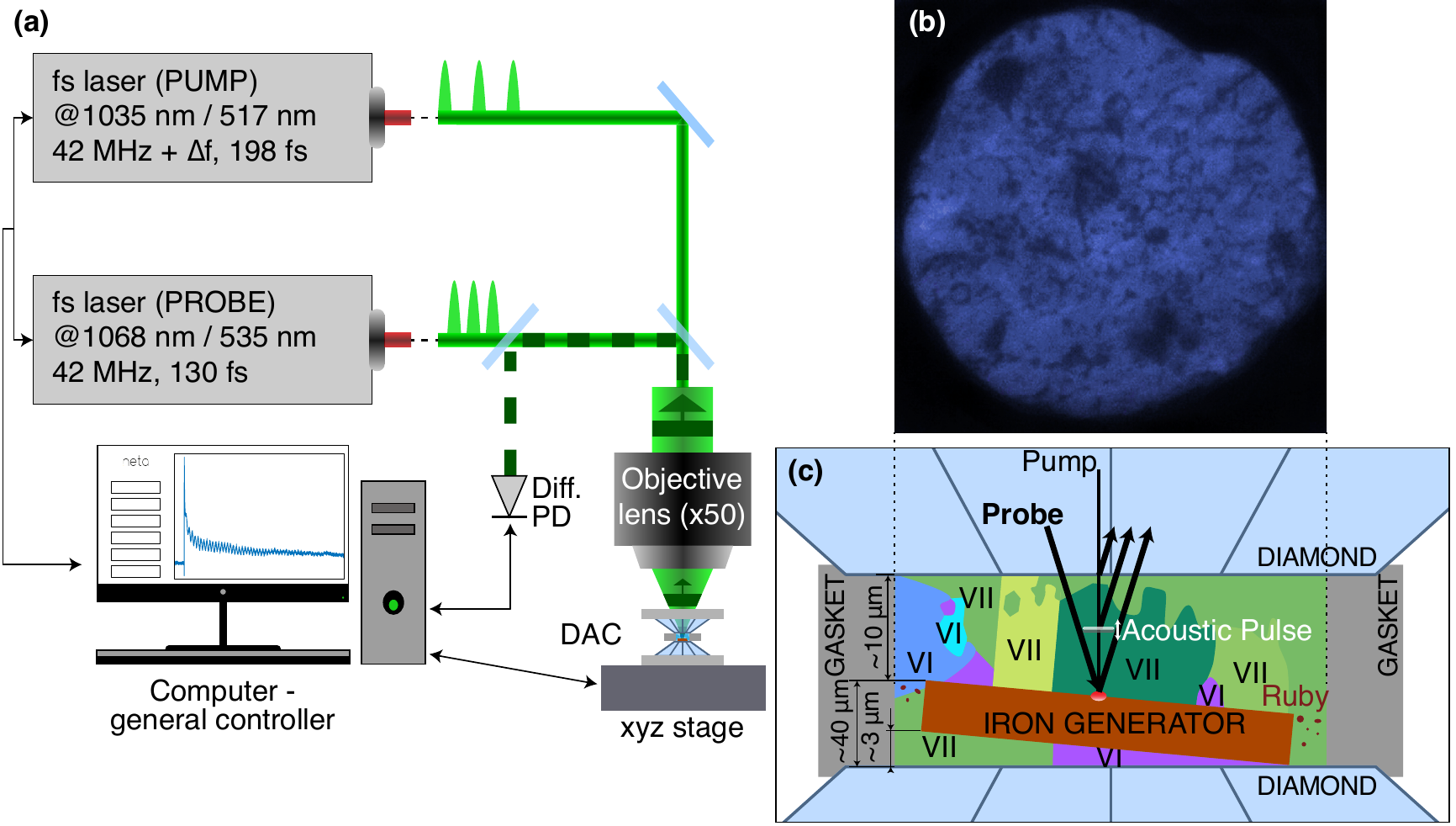}%[width=85mm]
\caption{(a) Experimental set-up of the picosecond acoustic microscope (JAX-M1, NETA, France) and (b) optical image of the \chemform{H_2O} ice polycrystal in the DAC between the diamond anvil and the iron optoacoustic transducer. (c) Magnified cross-sectional side view of the DAC. The results of the TDBS experiments (see Figs. 3 and 4) revealed the coexistence in the sample of two phases, ice VI (bluish) and ice VII (greenish). The disk-shape iron opto-acoustic generator inside the sample chamber has the diameter of about 110~\si{\micro\metre}. It touches the lower diamond anvil at its right end. Even though the pump and probe laser paths are collinear in the experiment, the probe one is shown inclined for a better visualisation of its different reflections.}
\label{fig:exp_setup}
\end{figure}

The experimental set-up is a commercial ASOPS-based picosecond acoustic microscope (JAX-M1, NETA, France) based on two pulsed fiber lasers radiating at the wavelengths of 517 and 535~\si{\nano\metre}, with the duration of pulses below 150~\si{\femto\second}, used as pump and probe, respectively [Fig.~\ref{fig:exp_setup}(a)]. We set the repetition rate of 42 MHz for the probe (leader cavity) and an offset repetition rate of 42.0005 MHz for the pump (follower cavity). More details on the set-up and the settings are given in the Methods section. The sample analysed in this work was a water ice polycrystal compressed in a DAC to a pressure of 2.15~\si{\giga\pascal} with an embedded iron optoacoustic transducer for launching the CAPs\cite{Kuriakose2017_v19_p053026}. The optical image of the \chemform{H_2O} ice polycrystal in the DAC shown in Fig.~\ref{fig:exp_setup}(b) reveals some dark areas on the iron transducer corresponding to low local optical reflectivity of iron, that could be due to some degradation processes\cite{Kuriakose2017_v19_p053026}.
%%%%%%%%%%%%%%%%%%%%%%%%%%%%%%%%%%%%%%%%%%%%%%%%%
%%%%%%%%%%%%%%%%%%%%%%%%%%%%%%%%%%%%%%%%%%%%%%%%%

\subsection*{Typical experimental signals}
In the TDBS experiments, the probe light scattered by moving CAPs and the probe light reflected by various stationary optical inhomogeneities of the sample is collected by a photodetector. In the considered experiments, the most important stationary sources of the probe light reflection are (i) the metallic optoacoustic transducer where the pump and probe laser pulses are co-focused and (ii) the interface between the diamond anvil---providing optical access to the compressed sample---and the water ice [see Fig.~\ref{fig:exp_setup}(c)]. Much weaker reflections are expected from the grain boundaries in case of the optically anisotropic grains of ice VI. Heterodyne detection in the TDBS technique is essentially interferometric and sensitive to the relative phase of the acoustically scattered and reflected probe light\cite{Thomsen1986_v60_p55--58, Grahn1989_v25_p2562--2569}. Because of a CAP propagation at its mode velocity, the phase of the light scattered by this CAP is continuously varying in time. The TDBS signal, due to the interferences on the photodetector of the acoustically-scattered probe light and the reflected one, is oscillating with extrema and zeros corresponding to the constructive and destructive interferences, respectively. The oscillating TDBS signal is commonly called the Brillouin oscillation (BO).

\begin{figure}[ht]
\centering
\includegraphics{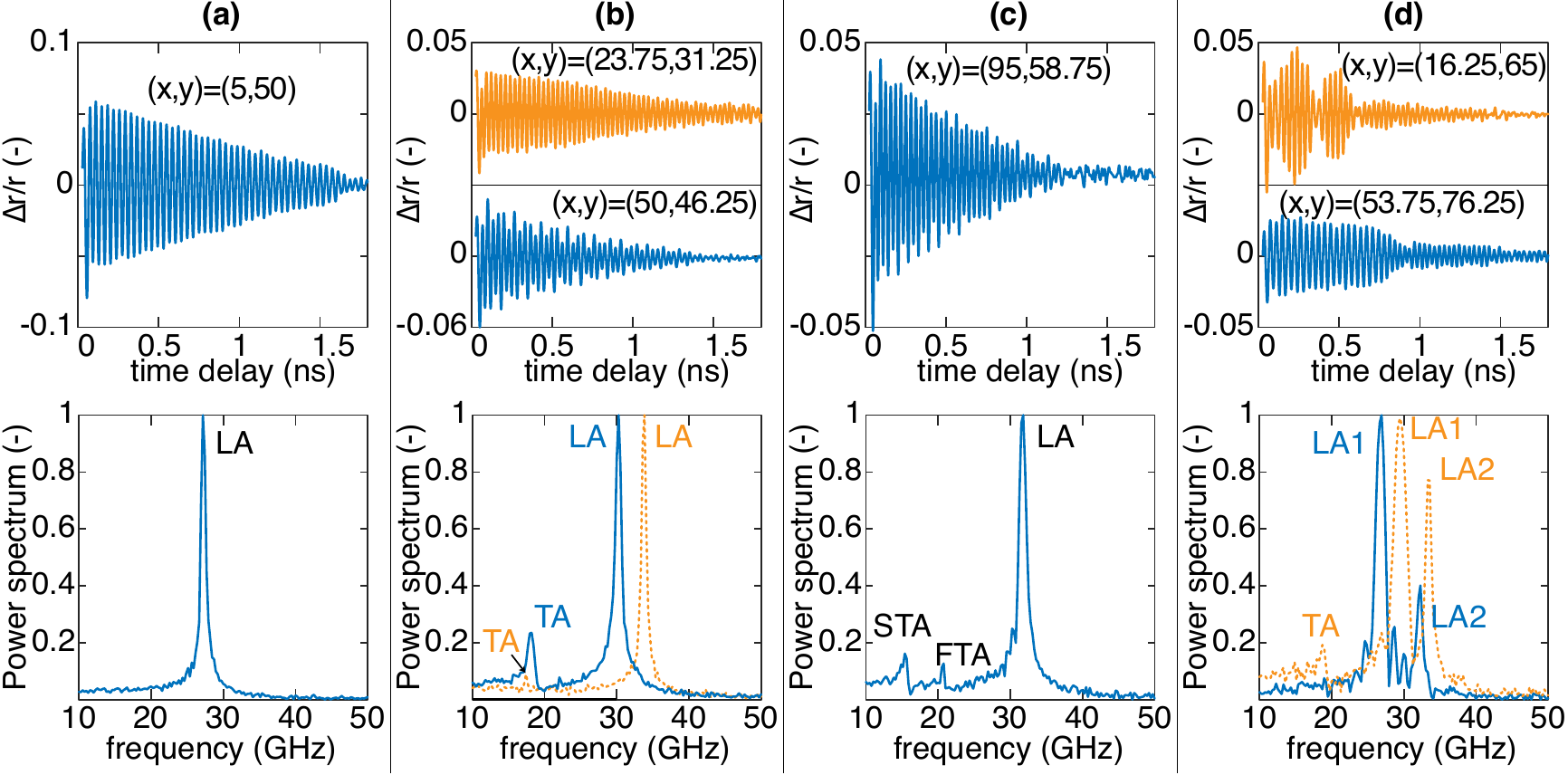}
\caption{Experimental acoustic contributions to transient reflectivity signals as a function of time delay (upper row) and their Fourier spectrum density (lower row) in four typical cases: (a) a single LA mode, (b) a LA and one TA modes with low (orange lines) and large (blue lines) relative shear amplitudes, (c) a LA and two TA modes, and (d) two LA modes propagating in two different grains \enquote{in parallel} (orange lines) and \enquote{in sequence} (blue lines). The coordinates x and y of the measurement points, which are indicated in the first row, correspond to the positioning (in \si{\micro\metre}) of the ice sample in the coordinate system presented in Figs. \ref{fig:2DBrillouinmap} and \ref{fig:3DSlices}.}
\label{fig:typical_signals}
\end{figure}

The typical oscillating TDBS signals, obtained (see Methods) for different lateral positions on the imaged area of the water ice sample, are presented in the first row of Fig.~\ref{fig:typical_signals}. The characteristic frequencies of the BOs in the TDBS signal correspond to the frequencies of the Stokes/anti-Stokes frequency shifts of the scattered light in the FDBS technique, \textit{i.e.}, to the Brillouin frequency (BF)\cite{Fabelinskii1968_v_p,Dil1982_v45_p285--334}. In our experimental geometry, where the coherent CAPs and the probe light are propagating collinearly, the most efficient process of the photon-phonon interaction is the backward scattering of the probe light. In this case, the BFs ($f_{B,\alpha}$) are related to the velocities ($\varv_\alpha$) of the coherent acoustic phonons, where $\alpha$ stands for the type of the acoustic mode (LA or one of the TA), by the following relation:
%\begin{linenomath*}
\begin{align}\label{eq:BrillouinFrequency}
    f_{B,\alpha} = \frac{2n\varv_\alpha}{\lambda_\text{probe}}\,,
\end{align}
%\end{linenomath*}
where $n$ is the refractive index of the transparent media at the wavelength in vacuum $\lambda_\text{probe}$ of the probe laser pulses. The Fourier analysis of the BOs is therefore commonly the first step used to reveal the physical origins of the detected TDBS signals.

Fourier spectra of typical BOs detected in our sample are presented in the second row of Fig.~\ref{fig:typical_signals}. These spectra clearly indicate that the TDBS signals have contributions from different types of the CAPs in the different lateral positions of our sample. This was used in the following for 3D TDBS imaging. Near the pressure of 2.15~\si{\giga\pascal}, corresponding to the transition at room temperature from the lower-pressure phase VI to the higher-pressure phase VII of \chemform{H_2O} ice\cite{Kuriakose2017_v19_p053026}, the available data on the optical and elastic properties (see Table~\ref{tab:properties}) provide the opportunity to estimate the expected BFs in both phases of ice. Please see the section Methods for the detailed explanations of these estimations and their uncertainties. The comparison of the spectra presented in Fig.~\ref{fig:typical_signals} with the results of these estimations, presented in the last row of Table~\ref{tab:properties}, provides confidence that not only LA modes but also TA modes are contributing to our detected TDBS signals for the first time in high-pressure TDBS experiments.
\begin{table}[ht]
\centering
\begin{tabular}{|l|c|c|}
\hline
 \makecell[l]{\textbf{Phases of \chemform{H_2O} ice} \\ (\textit{Lattice system})} & \makecell{\textbf{VI} \\ (\textit{Tetragonal})} & \makecell{\textbf{VII} \\ (\textit{Cubic})} \\
\hline
Density\cite{Polian1983_v27_p6409--6412, Shimizu1996_v53_p6107--6110} (\si{\gram\per\centi\metre\cubed}) & $\rho=1.419$ & $\rho=1.600$ \\
\hline
Elastic constants\cite{Shimizu1996_v53_p6107--6110} (\si{\giga\pascal}) & \makecell{$C_{11} = 40.56$ \\ $C_{12} = 13.75$ \\ $C_{13} = 18.59$ \\ $C_{33} = 34.43$ \\ $C_{44} = 7.50$ \\ $C_{66} = 6.39$} & \makecell{$C_{11} = 37.61$ \\ $C_{12} = 19.17$ \\ $C_{44} = 21.59$} \\
\hline
Refractive index @ 515 \si{\nano\metre}\cite{Polian1983_v27_p6409--6412, Shimizu1996_v53_p6107--6110} & $n=1.468$ & $n=1.521$ \\
\hline
Brillouin frequency (\si{\giga\hertz}) & \makecell{$f_{B,\text{LA}}\in\left[26.71-29.37\right]$ \\ $f_{B,\text{FTA}}\in\left[12.63-16.89\right]$ \\ $f_{B,\text{STA}}\in\left[11.66-12.80\right]$} & \makecell{$f_{B,\text{LA}}\in\left[27.58-33.09\right]$ \\ $f_{B,\text{FTA}}\in\left[16.46-20.90\right]$ \\ $f_{B,\text{STA}}\in\left[13.66-20.90\right]$} \\
\hline
\end{tabular}
\caption{\label{tab:properties}Properties of high-pressure phases of \chemform{H_2O} ice near the pressure 2.15~\si{\giga\pascal} and used to calculate the ranges of BFs for the LA, the fast TA (FTA) and the slow TA (STA) modes. The uncertainties in the estimated ranges of the Brillouin frequencies of different acoustic modes are discussed in Methods.}
\end{table}
The BOs presented in Fig.~\ref{fig:typical_signals}(a)-(c) can be attributed to the Brillouin scattering (BS) by a single LA mode, by a LA and one TA modes, and by a LA and two TA modes, \textit{i.e.} fast (FTA) and slow (STA), respectively. The BS by coherent transverse acoustic modes was earlier reported at ambient conditions in grains of \chemform{BiFeO_3}\cite{Lejman2014_v5_p4301,Lejman2016_v7_p12345} and of \chemform{CeO_2}\cite{Khafizov2016_v112_p209--215,Wang2019_v166_p34--38}. It is worth noting here that, potentially, in optically anisotropic ices, a single acoustic mode could produce up to three different BFs because of the birefringence phenomena giving rise to BS with optical mode conversion\cite{Lejman2014_v5_p4301,Lejman2016_v7_p12345}. However, the phase VII of \chemform{H_2O} ice is cubic (and thus optically isotropic), while the birefringence of the hexagonal phase VI is too weak (less than 1\%\cite{Polian1983_v27_p6409--6412}) to be observed in the here-reported TDBS experiments.
%%%%%%%%%%%%%%%%%%%%%%%%%%%%%%%%%%%%%%%%%%%%%%%%%
%%%%%%%%%%%%%%%%%%%%%%%%%%%%%%%%%%%%%%%%%%%%%%%%%

The beating in the BO evidenced in Fig.~\ref{fig:typical_signals}(d) demonstrates that the LA pulses are propagating in two different grains of ice. The orange lines correspond to their propagation in  both grains simultaneously, \textit{i.e.} \enquote{in parallel}. On the contrary, the blue lines correspond to the case when the LA pulses are propagating first in one grain and then in another one, \textit{i.e.} \enquote{in sequence}. The BOs corresponding to the sequential TDBS in two different media, \textit{i.e.} the TDBS imaging of the LA CAP transmission from one medium into another were already reported multiple times, for example for ice/diamond interface\cite{Nikitin2015_v5_p9352}, for the interface between ice VII and ice VI\cite{Kuriakose2017_v19_p053026} and for \chemform{SiO_2}/\chemform{Si} interface\cite{Devos2004_v70_p125208,Devos2005_v86_p211903}. The sequential propagation of the LA CAPs in two different grains of a polycrystal was reported as well\cite{Khafizov2016_v112_p209--215}. Note that there are also some reported observations of the simultaneous monitoring in two different media of two different LA CAPs propagating in the opposite directions: for example, the CAPs transmitted through and reflected from the interface between two media\cite{Devos2004_v70_p125208}. Here however, we report for the first time the TDBS monitoring of two LA CAPs propagating in parallel and in the same direction in two differently oriented ice grains along their mutual interface. In the next subsection, it is revealed that all the TDBS signals of the type of the orange one presented in Fig.~\ref{fig:typical_signals}(d) are detected in the vicinity of grain boundaries and are never detected in the grain volume. The latter suggests that the detection of these kind of TDBS signals could be fruitful for the localisation of grain boundaries, \textit{in situ} and in real time.

\subsection*{2D TDBS images obtained with LA and TA coherent acoustic pulses and grain identification}
\begin{figure}[ht]
\centering
\includegraphics{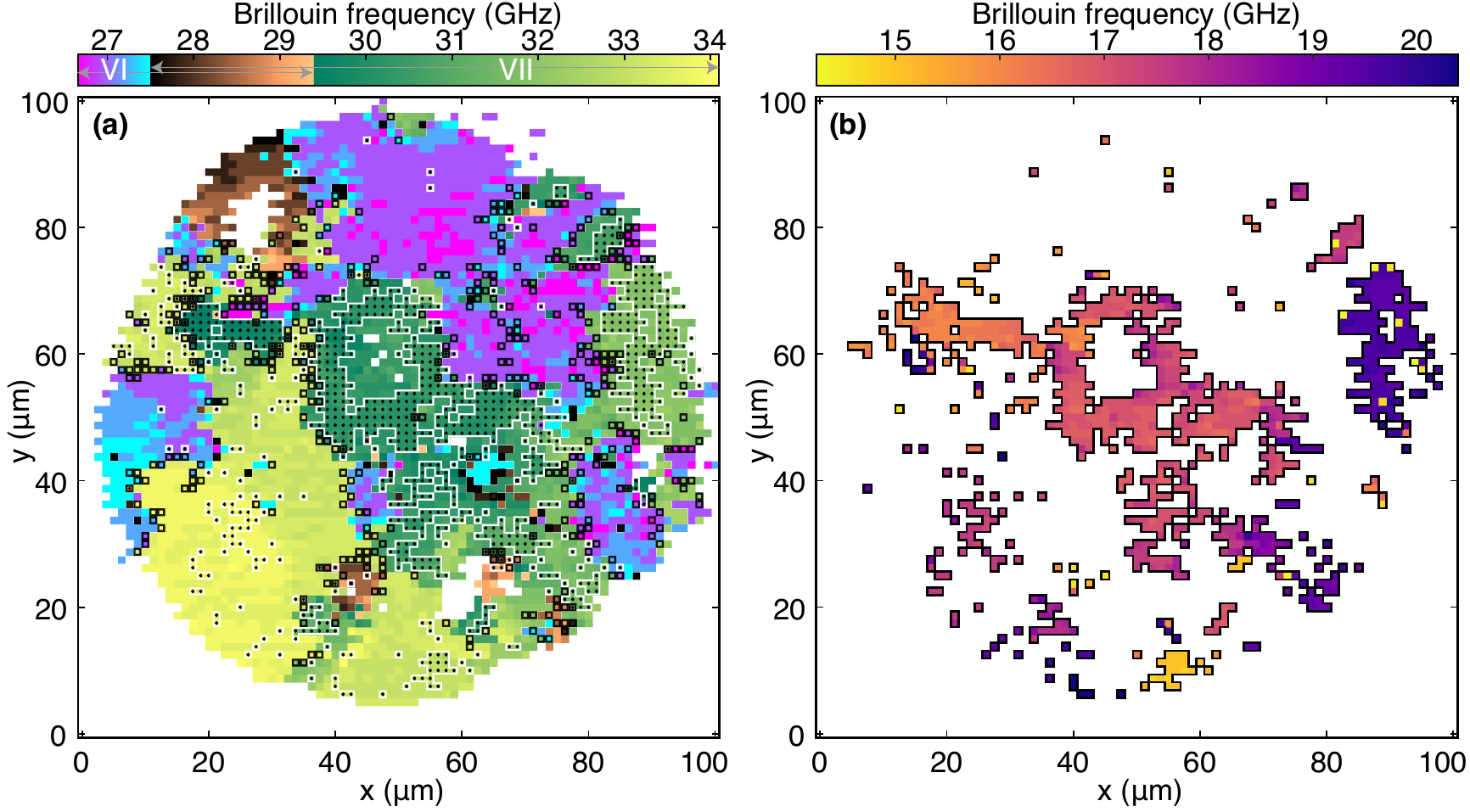}
\caption{Dominant frequency content attributed to (a) LA modes and (b) TA modes of the first two nanoseconds of the TDBS signals observed in the $100\times100$ \si{\micro\metre\squared} area of the ice sample. The white solid lines in part (a) delimit the areas where the TA modes shown in (b) have been detected. In (a), each pixel shown in (b) is marked with a central black dot. The black open square markers show the pixels where a beating phenomenon of the kind presented by the top (orange) signal in Fig.~\ref{fig:typical_signals}(d) has been observed.}
\label{fig:2DBrillouinmap}
\end{figure}

The color maps in Fig.~\ref{fig:2DBrillouinmap} represent the dominant frequency content attributed to (a) LA modes and (b) TA modes of the first two nanoseconds of the TDBS signals observed in the polycrystalline ice sample covering a round disk-shaped iron optoacoustic transducer of about 50~\si{\micro\metre} radius [Fig.~\ref{fig:exp_setup}(b,c)]. The chosen time interval corresponds to the acoustic propagation time through about 10 \si{\micro\metre} of the ice sample at a LA mode velocity. The TDBS signals are not observed in the locations indicated by white pixels due to poor signal-to-noise ratio. Note that the white pixel areas in Fig.~\ref{fig:2DBrillouinmap}(a) match the dark tones in the optical image of the sample [Fig.~\ref{fig:exp_setup}(b)]. Thus, it is expected that the low signals in these regions are due to the reduction in the probe light scattered by the iron optoacoustic transducer, which contributes to the heterodyning in TDBS detection. We used the prominence of the peak as a criterion to avoid the frequencies of the noise signal found in these locations. It is important to note that while the individual crystallites of the sample are not visible on the optical image, their presence is obvious in the TDBS imaging.

The black open squares in Fig.~\ref{fig:2DBrillouinmap}(a) depict the pixels where a beating phenomenon of the kind of the top (orange) signal in Fig.~\ref{fig:typical_signals}(d) has been observed. The automatic gathering of those pixels has been performed by tracking the signals which have two peaks in the spectrum density within the LA frequency range and which show at the same time a peak at the difference frequency in the spectrum density of its envelope. It is clearly visible from Fig.~\ref{fig:2DBrillouinmap}(a) that these signals are only found in locations in the vicinity of grain boundaries and never in a grain volume. We propose that this feature could be used for grain boundary imaging purposes (see Discussion).
%%%%%%%%%%%%%%%%%%%%%%%%%%%%%%%%%%%%%%%%%%%%%%%%%
%%%%%%%%%%%%%%%%%%%%%%%%%%%%%%%%%%%%%%%%%%%%%%%%%

To obtain the map of TA modes [(Fig.~\ref{fig:2DBrillouinmap}(b)], we have calculated prominent frequency peaks in the 10 to 22~\si{\giga\hertz} frequency range, which excludes the LA modes (see Table~\ref{tab:properties}). The TA map is less complete than the LA one because TA modes have usually lower amplitudes in TDBS signals than LA modes, making them more difficult to detect/observe. Nevertheless, even TA modes with weak amplitudes are detected in the data with good signal-to-noise ratio as shown by the dashed orange curve of the spectrum density in Fig.~\ref{fig:typical_signals}(b). The white solid lines in Fig.~\ref{fig:2DBrillouinmap}(a) indicate the areas where the TA modes shown in Fig.~\ref{fig:2DBrillouinmap}(b) have been detected. Each pixel shown in Fig.~\ref{fig:2DBrillouinmap}(b) is marked with a central black dot in Fig.~\ref{fig:2DBrillouinmap}(a). It is interesting to note that most of the detected TA modes are included in the zone where the frequency content is higher than 29.4~\si{\giga\hertz} (greenish color scale), showing that the TA modes with decent amplitudes are all found in the phase VII of \chemform{H_2O} ice, and mainly where the LA velocity is rather low (dark green) or high (light green). The lowest part of the color bar in Fig.~\ref{fig:2DBrillouinmap}(a) where the frequency is lower than 27.5~\si{\giga\hertz} (magenta-to-cyan color scale) depicts the crystallites belonging to the phase VI of \chemform{H_2O} ice. Interestingly, almost no TA modes are detected in these specific locations of the sample, which tends to indicate that the monitoring of TA modes in the phase VI is, for the present state of the technique, very unlikely, not to say impossible (see next section). Note that in the center-right grain (around $x\approx 90$~\si{\micro\metre} and $y\approx 60$~\si{\micro\metre}) in Fig.~\ref{fig:2DBrillouinmap}(b) with the lowest TA frequencies, the color of some neighbouring pixels is occasionally switching from dark violet, \textit{i.e.} the lowest frequency of the color scale, to light yellow, \textit{i.e.} the highest frequency of the color scale. This is because the figure represents the dominant frequency content in the TA frequency ranges and this particular grain contains two detectable TA modes. The amplitude of the slow (S) TA mode is usually larger than that of the fast (F) TA mode, except in some (light yellow) locations. The area where the identification of the main ice phase remains unclear from the first analysis of the dominant frequency content of the LA signals, \textit{i.e.}, where the frequency is greater than or equal to 27.5~\si{\giga\hertz} and lower than or equal to 29.4~\si{\giga\hertz} are depicted in a copper color in the scale in Fig.~\ref{fig:2DBrillouinmap}(a). 

\subsection*{3D TDBS imaging with LA coherent acoustic pulses}
%%%%%%%%%%%%%%%%%%%%%%%%%%%%%%%%%%%%%%%%%%%%%%%%%
\begin{figure}[hb!]
\centering
\includegraphics{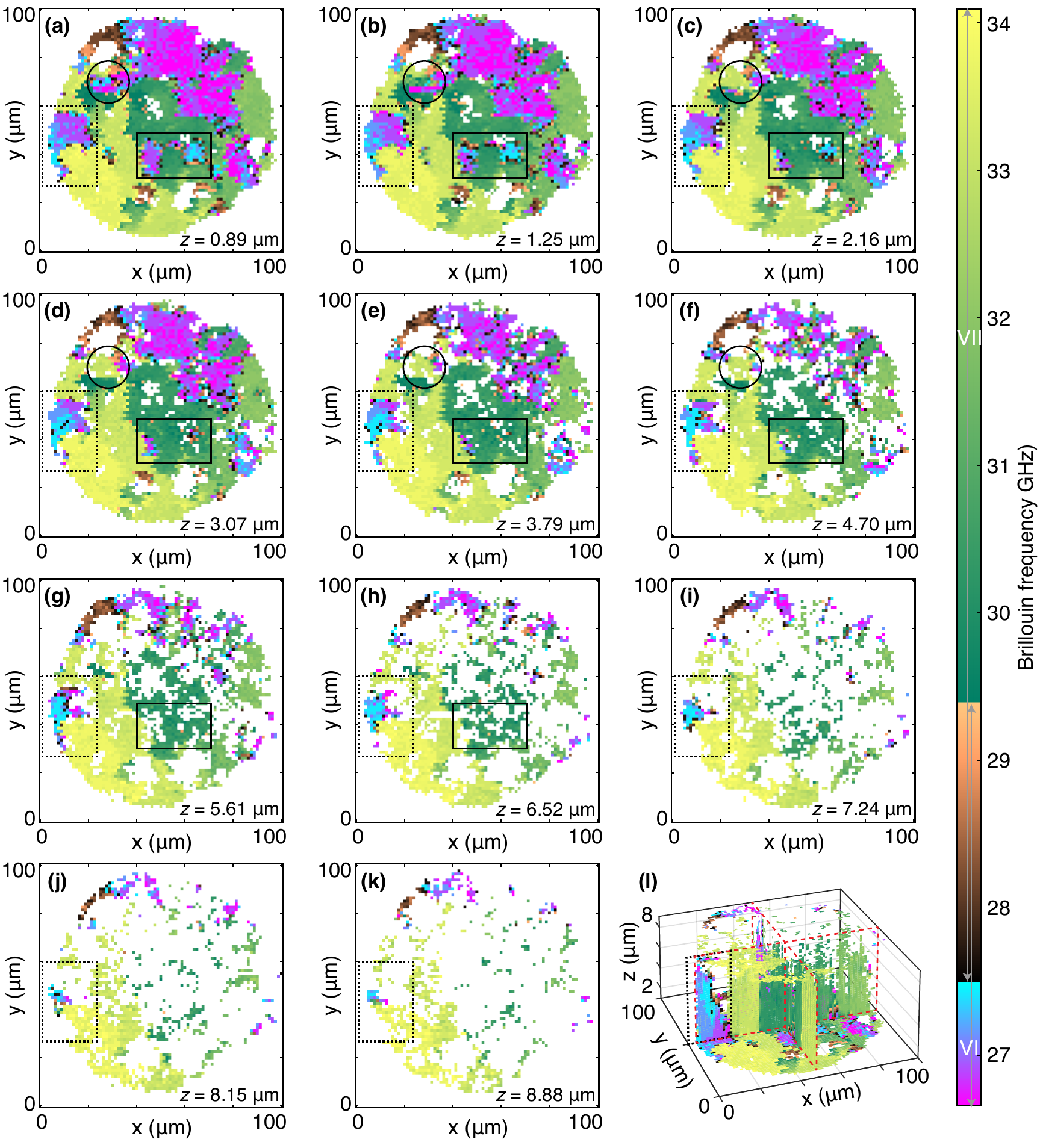}%[height=100mm]
\caption{3D TDBS imaging of the polycrystalline \chemform{H_2O} ice sample. (a)-(h) The slices are shown at particular positions along the $z$-axis shown in the right-bottom corner of each slice. (l) 3D representation of the complete probed volume with the first and last slices, (a) and (k) ($xy$ planes), and the middle slices ($yz$ and $xz$ planes) at $x=50$~\si{\micro\metre} and $y=50$~\si{\micro\metre}, respectively (red dashed rectangles).}
% \caption{3D TDBS imaging of the polycrystalline \chemform{H_2O} ice sample. The slices are shown at particular positions along the $z$-axis: (a) 0.89~\si{\micro\metre}, (b) 1.25~\si{\micro\metre}, (c) 2.16~\si{\micro\metre}, (d) 3.07~\si{\micro\metre}, (e) 3.79~\si{\micro\metre}, (f) 4.70~\si{\micro\metre}, (g) 5.61~\si{\micro\metre}, (h) 6.52~\si{\micro\metre}, (i) 7.24~\si{\micro\metre}, (j) 8.15~\si{\micro\metre}, and (h) 8.88~\si{\micro\metre}. (l) 3D representation of the complete probed volume with the first and last slices, (a) and (k) ($xy$ planes), and the middle slices ($yz$ and $xz$ planes) at $x=50$~\si{\micro\metre} and $y=50$~\si{\micro\metre}, respectively (red dashed rectangles).}
\label{fig:3DSlices}
\end{figure}
%%%%%%%%%%%%%%%%%%%%%%%%%%%%%%%%%%%%%%%%%%%%%%%%%
In opposition to the minor axial (depth) resolution of the classical FDBS technique, which would not be better than probe light wavelength even if confocal FDBS imaging is experimentally realised in a DAC, the TDBS technique could give access to a much higher axial resolution that could theoretically be equal to or even better than the wavelength of the acoustic mode scattering the light quanta of the used probe laser\cite{Gusev2018_v5_p031101,Steigerwald2009_v94_p111910,Mechri2009_v95_p091907,Lomonosov2012_v6_p1410--1415,Perez-Cota2016_v6_p39326}, \textit{i.e.}, twice shorter than the probe light wavelength. Reaching such resolution requires advanced time-frequency analysis of the TDBS signals which is out of the scope of this work. A usual short-time Fourier transform is here used to demonstrate capability of the 3D imaging with the TDBS method at high pressure. Fig.~\ref{fig:3DSlices} represents the depth-resolved information of the ice sample state recovered using the LA mode. Although the same imaging with TA coherent acoustic pulses is possible, the current detected TA modes are too sparse [see Fig.\ref{fig:2DBrillouinmap}(b)] to report here on the possibility to obtain an informative 3D image.

In Fig.~\ref{fig:3DSlices}, we estimate the dominant LA-mode-related frequency at different depths by calculating the FFT of the acoustic signal sliced with a Hann window of 0.23~\si{\nano\second} (about 7 oscillations of the LA mode), which gives a depth resolution of about 1.2~\si{\micro\metre}. Using the temporal indication of the center of the sliding window and the measured local velocities at the previous instants for the corresponding acoustic mode, the time axis can be transformed in the depth axis; the smaller the time, the closer the probed zone to the iron optoacoustic transducer. Note that this change of coordinates implies knowledge of the local refractive index. For the sake of simplicity, we have chosen here to attribute only one refractive index to a given pixel, even though it contains two phases. The attribution of the refractive index is based on the previous determination of the phases of each grain as discussed in Fig.~\ref{fig:2DBrillouinmap}(a). This means that we used the refractive index of the dominant phase. We are aware that this will engender some distortions of the grains, which we consider as unimportant in this qualitative analysis. For pixels with unattributed phases ($f_{B,\text{LA}}\in\left[27.5,29.4\right]$~\si{\giga\hertz}), the refractive index is chosen to be the average of that of phases VI and VII. The color map is common for all parts of Fig.~\ref{fig:3DSlices} and is the same as in Fig.~\ref{fig:2DBrillouinmap}(a). Each part of the figure, except the last one (l), is a slice of the polycrystalline \chemform{H_2O} ice sample at a particular position along the $z$-axis shown in the right-bottom corner of each slice. In Fig.~\ref{fig:3DSlices}(l), a 3D representation of the complete probed volume is shown with the first and last slices, (a) and (k) ($xy$ planes), and the middle slices ($yz$ and $xz$ planes) at $x=50$~\si{\micro\metre} and $y=50$~\si{\micro\metre}, respectively (red dashed rectangles). The rectangles (shown with solid and dotted lines) and the circle mark in Fig.~\ref{fig:3DSlices} indicate the sample regions/volumes with some interesting changes in the bulk revealed by this 3D imaging which are discussed in the following.

From the 2D analysis of Fig.~\ref{fig:2DBrillouinmap}(a), the volume zone pointed out by the circle in Fig.~\ref{fig:3DSlices}(a)-(f) is expected to be composed of multiple crystallites since it contains numerous TDBS signals with beatings. It can indeed be seen in Fig.~\ref{fig:3DSlices} that close to the iron optoacoustic transducer, this zone is made of multiple and relatively small crystallites depicted by the different colors. About half of the pixels within the circle in Fig.~\ref{fig:3DSlices}(a) belong to an ice VI crystallite. In the next slice at $z=1.25$~\si{\micro\metre}, Fig.~\ref{fig:3DSlices}(b), it can be seen that in the place of the crystallites of phase VI with high LA velocity (cyan, $f\approx 27.5$~\si{\giga\hertz}) the phase is now VII since the pixels are green. From that slice to the next one at $z=2.16$~\si{\micro\metre}, Fig.~\ref{fig:3DSlices}(c), the color slightly changes to lighter green, which could be temptatively attributed to a change from one phase VII crystallite to another. Yet, the resolution along $z$ of about 1.2~\si{\micro\metre} controlled by the width of the Hann window means that the frequency/color attributed to those pixels in the slice in Fig.~\ref{fig:3DSlices}(b) is indeed a weighted average of the frequency of the phase VI crystallite (close to the transducer) and that of the phase VII [higher in the DAC, see Fig.~\ref{fig:exp_setup}(c) for illustration]. A precise depth localisation of the boundary between the two could be done but requires a more sophisticated signal processing. The rectangles shown by solid lines [Fig.~\ref{fig:3DSlices}(a)-(h)] depict also the zones where crystallites of the phase VI are seen between the optoacoustic transducer and the higher located crystallites of the phase VII. Interestingly, the color is not switching here from that of phase VI to that of phase VII at the same depth for all pixels. This is the clear sign of an oblique interface between the two phases of \chemform{H_2O} ice. Last but not least, in the rectangles shown with dotted lines [Fig.~\ref{fig:3DSlices}(a)-(l)] we could recognise another clear example of the 3D imaging capability of the TDBS technique. In this case, oblique boundaries between several crystallites of phase VI having different orientations are imaged in 3D and up to the ice/diamond interface. In Fig.~\ref{fig:3DSlices}(l), thanks to the middle slice ($xz$ plane) at $y=50$~\si{\micro\metre}, the inclined boundaries are clearly visible. It is important to note that for the sake of visibility of that 3D figure, the ratio of the $z$-axis length to that of the $x$- and the $y$-axes is set to 4.

\section*{Discussion}

% \subsection*{On the TDBS depth spatial resolution}
The 3D TDBS image shown in Fig.~\ref{fig:3DSlices} and obtained with the LA modes has a different depth resolution for each pixel since it depends on the local acoustic wavelength, and is overall of about 1.2~\si{\micro\metre}, as mentioned previously. The lateral resolution of the image is here fixed by the focal spot size of the laser beams: 2.5~\si{\micro\metre}, which is approximately twice larger than the chosen lateral step of the scan. Here, we have chosen to use the time-frequency analysis technique known as the short-time Fourier transform. The drawback of this technique is the trade-off one should choose between the frequency resolution and the temporal resolution (\textit{i.e.} depth resolution). To have a chance to differentiate frequency changes in the interval of the LA Brillouin frequency of phase VI that spans only over 2.7~\si{\giga\hertz} (see Table~\ref{tab:properties} and Fig.~\ref{fig:3DSlices}), the temporal size of the Hann window has been chosen to be 0.23~\si{\nano\second}, which is theoretically even a bit small since the 3 dB bandwidth of such a window function is of 3.2~\si{\giga\hertz}. With a better-suited signal processing method (e.g. temporal fitting with a model signal), the axial resolution could be improved to about 170~\si{\nano\metre} with both LA mode and TA modes since it is only limited by the probed acoustic wavelength $\lambda_\text{B}=\frac{\lambda_\text{probe}}{2n}$.

% \subsection*{On the TDBS depth of imaging}
In the TDBS imaging experiments, the depth of imaging could be limited by several factors such as the coherence length of the probe laser pulses in the medium under evaluation, the diffraction length of the probe laser radiation and of the coherent acoustic waves, as well as by the absorption of the acoustic pulses and probe light. For the probe laser pulses of the duration $\tau_\text{probe}\approx 150$~\si{\femto\second} and the refractive index of ice at 2.15~\si{\giga\pascal} of $n\approx 1.5$ at the vacuum optical wavelength of  515~\si{\nano\metre}\cite{Polian1983_v27_p6409--6412,Shimizu1996_v53_p6107--6110} (close to our probe wavelength $\lambda_\text{probe}\cong535$~\si{\nano\metre}), the estimate of the coherence length  $L_\text{coherence}^\text{probe}\equiv c_0 \tau_\text{probe}/(2n)$, where $c_0$  denotes the speed of light in vacuum, is $L_\text{coherence}^\text{probe}\approx 15$~\si{\micro\metre}. Because of the strict relation between the probe optical wavelength in the medium and the wavelength of the coherent acoustic phonon at the Brillouin frequency, \textit{i.e.} $\lambda_B=\frac{\lambda_\text{probe}}{2n}$, the Rayleigh range (\textit{i.e.}, the diffraction length) of both the probe light beam and the coherent acoustic beam at Brillouin frequency, can be described by the same formula: $L_R\equiv\frac{\pi a^2}{(\lambda_\text{probe}/n)}$, where $a$ is the radius at $1/e^2$  level of the intensity distribution in the probe or in the pump laser beams. The focusing of the pump laser beam controls the radius of the photo-generated coherent acoustic pulse. In our experiments, the pump and probe laser beams are co-focused in the same spot with a radius $a\approx 1.25$~\si{\micro\metre}. Thus, the diffraction length in our experiments is about $L_R\approx 14.3$~\si{\micro\metre}, which is very similar to the coherence length. However, as the signal amplitude falls down $e^2$ times at the coherence length, while it falls down only $\sqrt{2}$ times at the Rayleigh range, the influence of the diffraction ($\sim 1/[1+\left(\varv_\text{LA} t/L_R\right)^2]^{1/2}$) on the depth of imaging is expected to be negligible in our experiments.

Experimentally, in some lateral positions of the sample, where the ice between the optoacoustic transducer and the diamond anvil was a single crystal, we observed Brillouin oscillations through the complete distance between the iron generator and the ice/diamond interface [see Fig.~\ref{fig:exp_setup}(c) and Supplementary information]. In these positions, the velocity of the quasi-longitudinal acoustic wave at Brillouin frequency of 26.85~\si{\giga\hertz} is estimated to be $\varv_\text{LA}\approx 4881$~\si{\metre\per\second}. This wave crossed the distance between the optoacoustic generator and the diamond anvil in 2.15~\si{\nano\second}, providing the estimate of the distance between them in a particular lateral position as 10.5~\si{\micro\metre}. Note that, at this propagation distance, the Brillouin oscillation exhibited an amplitude reduction of only 75\%. We fitted the observed Brillouin amplitude decay by accounting for the Gaussian temporal decay related to the coherence length of the probe laser pulses and of the additional exponential temporal decay that could be potentially caused by unknown absorption of coherent acoustic waves. The model signal used for the fit reads: $A\exp(-\alpha t)\exp[-2(t/\tau)^2]\cos(2\pi f_{B,\text{LA}} t + \phi)$, with $A$ the amplitude, $\alpha$ the acoustic absorption coefficient, $\tau$ the coherence time of the probe laser pulses and $\phi$ the phase of the signal (see Supplementary information for details). The absorption of probe green light is known to be negligible at the evaluated coherence length. These fits demonstrated that the contribution of the acoustic absorption in the observed decay is negligible as well. The decay of the TDBS signal in single crystals of ice is controlled by the coherence length, while the depth of imaging is limited by the sample thickness. The fitting procedure revealed the coherence length of around 13~\si{\micro\metre} and the duration of the probe laser pulses about 125~\si{\femto\second}, which is consistent with the pulse duration of the probe laser.

% \subsection*{On the TDBS with STA and FTA CAPs}
Even though the TA modes showed up too sparsely to provide a qualitative 3D image as it has been done in Fig.~\ref{fig:3DSlices} with the LA mode, the TA modes [Fig.~\ref{fig:2DBrillouinmap}(b)] are of a great importance and complementary to the LA mode to identify the unknown grains and to estimate a set of possible orientations of each crystallite. The latter was done from the estimation of the propagation directions of the coherent acoustic pulses in which the LA and TA modes would lead to the measured Brillouin frequencies. The TA modes could even be the only way to distinguish two grains having different orientations but the same LA velocity. The contrast in the TA-mode-based images could be stronger than in LA images in the case of a stronger anisotropy of the TA modes and could therefore give additional insight on the sample state\cite{Wang2019_v166_p34--38}. In our case, the use of the information about the presence or not of the TA mode, together with the LA-mode-based 3D image, helps to recognise to which phase of \chemform{H_2O} ice belongs the related grains.

% \subsection*{On detection of the TA modes}
Detection of the TA modes depends on numerous parameters. First, the amplitude of the signal, \textit{i.e.}, of the photoelastic interaction between the probe laser photon and the TA coherent phonon, depends in general on the amplitude of the acoustic strains a given TA mode engenders, of the probe laser pulse power and, at the same time, of the probe laser polarisation\cite{Lejman2016_v7_p12345,Wang2020_v11_p1597}. The power of the probe light is commonly optimised simultaneously with the power of the pump light for achieving the maximum TDBS signal amplitudes without introducing irreversible degradation of the optoacoustic generator and changes in the state of the sample. Optimisation of the probe light polarisation or detection of the depolarised reflected and acoustically-scattered probe light enhances the amplitudes of the detected TA modes for particular orientations of crystallites--yet this was not performed in this work. Second, the amplitudes, of the overall TDBS signal and of each individual acoustic mode contributing to it, depend on the generation of the CAPs. The generation of the LA and both TA modes is achieved here by the mode conversion of the plane LA mode thermoelastically-generated in the isotropic optoacoustic transducer by the absorption of the pump laser pulses. The thermoelastic generation of the CAPs directly in water ice caused by the heat transfer from light absorbing iron to transparent ice is estimated to be negligible. If the interface between the optoacoustic transducer and the \chemform{H_2O} ice is flat and perfect, then the elastic forces (stresses) and the mechanical displacements across the interface are continuous. In that case, the generation efficiency of each acoustic mode in the \chemform{H_2O} ice is directly proportional to the efficiency of the conversion, \textit{i.e.}, of the transfer, of the acoustic energy from the excited mode(s) in the transducer to each acoustic mode in the ice that are expected for the particular orientation of the crystallite relative to the interface. In our case, this could be solved by calculating the acoustic transmission coefficients, from an isotropic medium (the transducer) to an anisotropic medium (an arbitrary-oriented crystallite of the polycrystalline \chemform{H_2O} ice), of a plane LA wave transmitted and mode-converted at the interface to a plane LA and two plane TA modes, respectively. Since the lateral size of the pump laser focus spot is one order of magnitude larger than the probed acoustic wavelength, this transmission problem could be solved as one dimensional for the plane acoustic modes propagating normally to the interface. While the relative orientation of the principal axis of the crystallite changes with respect to the normal to the transducer/ice interface, so do the transmission coefficients and thus the generation efficiency of each acoustic mode. Finally, to reveal the possible reasons of the TA mode absence/presence in our TDBS signals, we assume that the detection of the different acoustic modes predominantly takes place via the interaction of the probe light with the longitudinal, \textit{i.e.}, the $z$ strain component, of the acoustic modes.

\begin{figure}[htb!]
\centering
\includegraphics{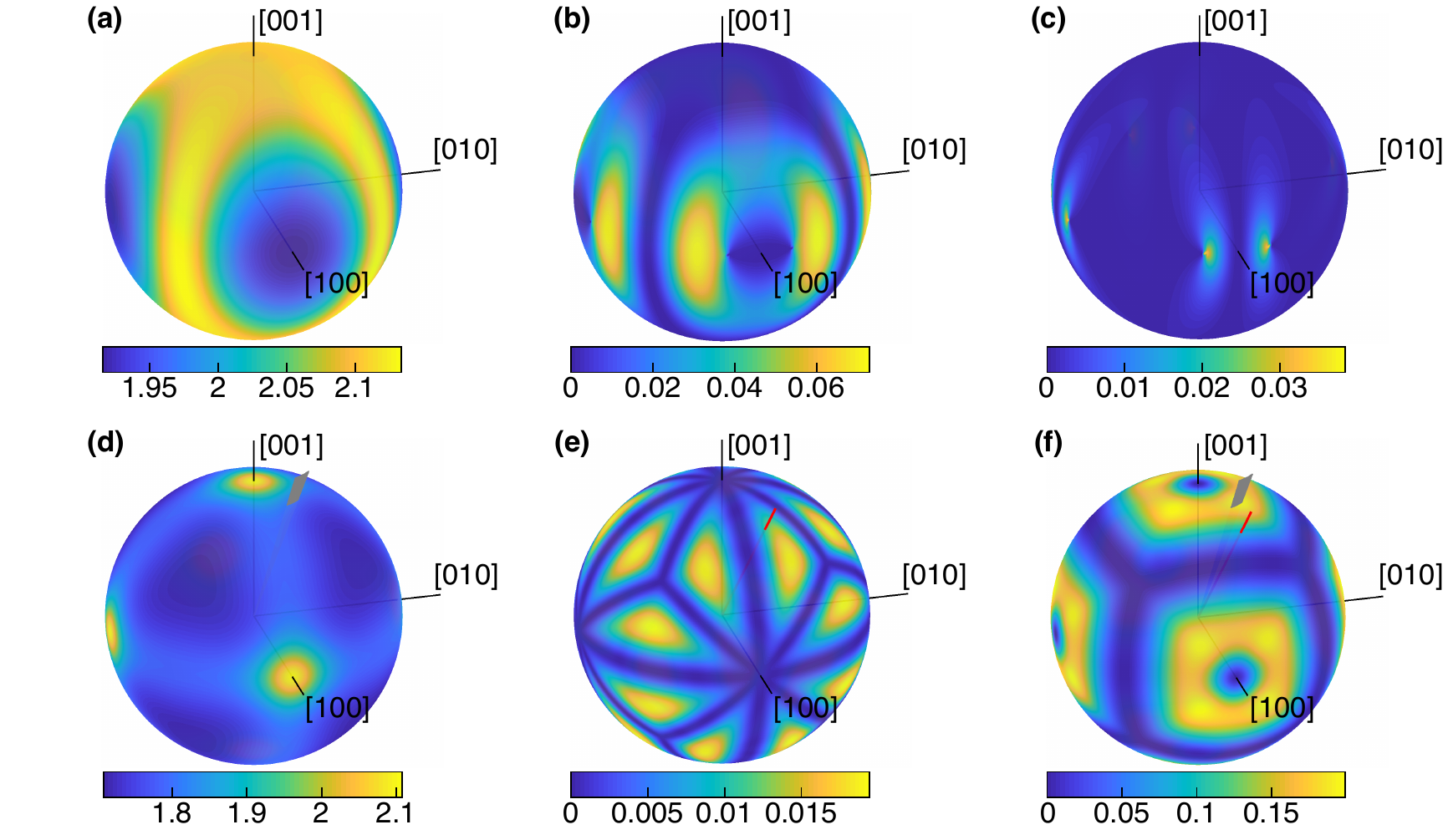}%
\caption{Generation efficiency of each acoustic mode for (top) the phase VI and (bottom) the phase VII of \chemform{H_2O} ice along the principal axes of the crystals as a function of the orientation of the normal to the iron/ice interface relative to the ice crystal principal axes. The results are presented separately for different modes: in (a,d) for the LA mode, in (b,e) for the FTA mode, and in (c,f) for the STA mode.}
\label{fig:GeneEff}
\end{figure}

% \subsection*{On the application of the detected quasi-shear TDBS signals}
Considering the mechanical properties of both phases of \chemform{H_2O} ice reported in Tab.~\ref{tab:properties} and that of the cast iron (isotropic) to be $\rho_\mathrm{Fe}=7.87$~\si{\gram\per\centi\metre\cubed}, $C_{11}^\mathrm{Fe}=279$~\si{\giga\pascal}, and $C_{44}^\mathrm{Fe}=82$~\si{\giga\pascal}, the generation (transmission and mode conversion) efficiencies of each acoustic mode is plotted in Fig.~\ref{fig:GeneEff} for (top) the phase VI and (bottom) the phase VII of \chemform{H_2O} ice. The surfaces in Fig.~\ref{fig:GeneEff} are plotted along the principal axes of the crystal as a function of the orientation of the $z$-axis (the normal to the iron/ice interface, \textit{i.e.} the propagation direction), relative to the ice crystal principal axes. The results are presented separately for different modes: in (a,d) for the LA mode, in (b,e) for the FTA mode, and in (c,f) for the STA mode. The transmission/mode conversion coefficient in Fig.~\ref{fig:GeneEff} are defined as the ratios of the longitudinal, i.e. of the $z$ component, in the generated strains associated to the acoustic eigenmodes of ices to the longitudinal strain of the acoustic wave in iron incident on the iron/ice interface. First, it is noticeable for both phases of ice that the magnitudes of the generation efficiency are one or two orders of magnitude larger for the LA mode in any direction than for both TA modes. Note also that the generation efficiency of the STA mode in phase VII is one order of magnitude higher than that of any other TA mode, which therefore partly explains why TA modes have not been observed for the phase VI and why mostly a single TA mode has been observed for the phase VII. According to our calculation, it is very probable that in all measurements with only one TA mode detected in phase VII the STA mode was observed. Careful analysis of the pixels of unidentified phases (copper color scale) in Fig.~\ref{fig:2DBrillouinmap}(a) suggested that some black dots are present in the two center-bottom grains, while the top-left one is completely free of them. Reminding that a black dot denotes the presence of TA modes and in the view of the above-discussed likelihood to have large enough TA mode amplitudes, we believe that it is safe to assign both center-bottom grains to the phase VII while the top-left one to the phase VI. Within a global view of the $xy$ distribution of both phases, it is also seen that the center-bottom grains are surrounded by the phase VII whereas the top-left one is surrounded by the phase VI, which supports our analysis. Thus, in our sample the experimental results on the observed TA modes can be qualitatively explained under the assumption that their detection takes place via scattering of the probe light by their longitudinal strain components, without accounting for the scattering of the probe light by their shear strain components.

In Fig.~\ref{fig:2DBrillouinmap}(b), we have discussed the case of the grain located at around $x\approx 90$~\si{\micro\metre} and $y\approx 60$~\si{\micro\metre} of phase VII which shows two TA modes. It was pointed out that the dominant TA frequency content was usually the lowest of the two TA modes, \textit{i.e.} the STA mode, which is consistent with our calculation. The detection of the three acoustic modes could be used to find out the orientation of the CAP propagation direction in the principal axes of the grain. Knowing that in that grain the Brillouin frequencies are $\sim31.8$~\si{\giga\hertz}, $\sim20.8$~\si{\giga\hertz}, and $\sim15.4$~\si{\giga\hertz}, the propagation direction forms an angle $\varphi$ of \ang{17}-\ang{18} and an angle $\theta$ of \ang{32}-\ang{33}, where $\varphi$ and $\theta$ are the proper Euler angles of intrinsic rotations about moving axes $e_Z$-$e'_X$-$e''_Z$\cite{Nolze2015_v50_p188-201}. Of course, all other directions symmetry-equivalent to the evaluated direction are also possible. If we report that direction in Fig.~\ref{fig:GeneEff}(e)-(f) (red thick line), it is seen that it corresponds to a direction in which the FTA mode has a generation efficiency along the $z$-axis of about 0.01 (half of the highest possible efficiency) and that of the STA mode is of about 0.085, which explains the ease in the simultaneous detection of these two modes. If only one TA mode is measured, as in the center of the sample made of phase VII, it is likely that this mode is the slow one and the orientation of the $z$-axis in the crystal principal axes is not unique anymore but defines a part of a cone ($\theta=\ang{20}$, $\varphi\in[\ang{26},\ang{45}]$) shown in grey in Fig.~\ref{fig:GeneEff}(d,f). This is the case of the TDBS signal in Fig.~\ref{fig:typical_signals}(b) at the position $(x,y)=(50,46.25)$~\si{\micro\metre}. Obviously, all other symmetric directions of that partial cone in the cubic crystal are also possible. If only the LA mode is detected as in almost all TDBS signals attributed to phase VI of \chemform{H_2O} ice, the possible range of orientations of the $z$-axis in the crystal principal axes is too uncertain for description here which makes any valuable conclusion impossible. Importantly, the detection of the LA mode and of one of the TA modes, which type, i.e. STA or FTA, is known (here from the analysis of the generation efficiency) allows to reduce the range of orientations to limits, which could already be of interest for analysing polycrystalline materials and the evolution of the polycrystal upon change of temperature and/or pressure. However, it is worth noting here that the determination of the propagation direction in a crystallite does not mean the complete orientation of the crystallite relative to other crystallites, because the crystallite could be arbitrary rotated around the $z$ axis (interface normal). The complete orientation of the crystallites could be obtained by performing TDBS measurements with different polarisations of the probe light\cite{Wang2020_v11_p1597}.

% \subsection*{On the use of the detected TDBS signals with beatings}
Last but not least, we propose that the detection of particular TDBS signals, corresponding to the simultaneous propagation of LA pulses in two adjacent grains as depicted by the black open squares in Fig.~\ref{fig:2DBrillouinmap}(a), provides an original opportunity to localise the boundaries between the grains. If the interest of an experiment is only in imaging the grain boundaries, this observation could be used to follow grain boundaries and their movement, if any, in real time while performing the 3D TDBS imaging. To do so, a feedback signal could indeed be designed out of the observation of beatings to control the motorised stage and hence to limit the number of measurement points, which would accelerate the imaging procedure of grain boundaries compared to a full scan of the sample. The first step of such experiment of grain boundary imaging would be to find a first location where beatings are visible in the TDBS signal. Then, around that point and with a spatial step that could be lower than the laser spot diameters, an automatic optimisation of the sample positioning allowing to maximise the beating phenomenon could be done, which would give the direction to follow the grain boundary. By repeating this process step by step, an image of grain boundaries could be obtained in a much shorter time than the image obtained from a full scan of the specimen.

\section*{Conclusions}
We reported on advances in applications of the TDBS technique in high-pressure experiments in a DAC that followed the recent achievements of the 3D TDBS imaging at ambient conditions. The results of the first high-pressure 3D TDBS imaging experiments and of the first observation in a DAC of the TDBS with quasi-shear CAPs were presented. We demonstrated some examples of the fruitful application of the TDBS imaging with several acoustic modes simultaneously. We also revealed the possibility to localise positions of grain boundaries in polycrystal by the identification of the specific TDBS signals that are due to the simultaneous propagation of an acoustic pulse in two adjacent grains. Overall our reported results are a big step towards the perspective of the full 3D characterisation of sample texture at extreme conditions (high pressures and/or temperatures) and its evolution on further compression or temperature change. Such characterisation is of special interest for different branches of research at extreme conditions. It will make possible an examination of the texture of minerals present in the deep Earth and its evolution upon nonhydrostatic compression with the detailedness presently not accessible by other techniques. Such information will permit conclusion about the nature of seismic anisotropies observed in the Earth's mantle. Also, it could help to quantitatively investigate kinetics of phase transitions and chemical reactions at high pressures and temperatures as well as relation between the crystallographic orientations of crystallites of HP-HT phases with respect to consumed crystallites of lower pressure phases. Obviously, further improvement of the signal treatment and resolution of the TDBS 3D imaging permitting recovery of quantitative values of shear sound velocities will further extend the horizons in investigation of solids at extreme conditions.

\section*{Methods}

\subsection*{DAC and the sample}
The high-pressure experiments on water ice were performed using a diamond anvil cell (DAC) of the Merrill-Bassett type\cite{Merrill1974_v45_p290-294}. We used diamond anvils with unbevelled culets having a diameter of $\sim500$~\si{\micro\metre}. A hole of $\sim180$~\si{\micro\metre} in diameter drilled in the center of a stainless steel gasket pre-indented to the thickness of $\sim60$~\si{\micro\metre} represented the sample volume [Fig.~\ref{fig:exp_setup}(c)]. The latter was filled with bi-distilled water that solidified upon compression. The sample volume contained also a thin iron disk and few ruby grains of 1-5~\si{\micro\metre} in size distributed in the space between the iron disk and the gasket wall. The iron disk served as the opto-acoustic generator for launching coherent acoustic pulses into the ice sample. It was obtained by a gentle compression of a small iron spherule between the diamond anvils, which surfaces are parallel to each other, until the desired thickness was obtained. The disk was initially in contact with one of the anvils but lifted up on one side by a few microns when the sample volume was filled with water [Fig.~\ref{fig:exp_setup}(c)]. Note that the shape and the orientation of the iron generator presented in Fig.~\ref{fig:exp_setup}(c) are illustrative. The ruby grains were used to measure pressure using the calibrated shift of the R1 fluorescence line with compression\cite{Mao1986_v91_p4673-4676}. Finally, the \chemform{H_2O} ice sample was compressed to 2.15~\si{\giga\pascal} until the ice VI and ice VII phases were simultaneously present in the sample volume. The fluorescence spectra were collected with an Ocean Optics USB4000 spectrometer. Within the instrument resolution, the pressure was found to be the same in all points around the sample. The collected spectra were fitted with the Gaussian-Lorentzian function to obtain position of the fluorescence line maximum. Corresponding pressure was calculated to be $2.15\pm 0.05$~\si{\giga\pascal} using the calibration curve for quasihydrostatic load conditions\cite{Mao1986_v91_p4673-4676}.

\subsection*{Pump/probe ASOPS set-up}
The TDBS imaging was performed using a picosecond acoustic microscope (JAX-M1, NETA, France)\cite{Dilhaire2010_v_p} with pump and probe laser beams following the same path within the DAC, as shown schematically in Fig.~\ref{fig:exp_setup}(a). Two pulsed fiber lasers of optical wavelength of 1034.8~\si{\nano\metre} and 1068.4~\si{\nano\metre}, of pulse duration 198~\si{\femto\second} and 130~\si{\femto\second}, respectively, and with a repetition rate of 42~\si{\mega\hertz} are synchronised for asynchronous optical sampling\cite{Bartels2007_v78_p035107}. The repetition rate of the follower laser cavity is slightly offset compared to that of the leader one. The used offset in our measurements of 500~\si{\hertz} corresponds to a temporal sampling of 0.28~\si{\pico\second}. The two optical wavelengths obtained by frequency doubling of the fundamental radiations, 517~\si{\nano\metre} and 535~\si{\nano\metre}, were used as pump (follower cavity) and probe (leader cavity), respectively. The beams were normally incident and co-focused to the surface of the iron disk surface to the spot of approximately 1.25~\si{\micro\metre} radius at $1/e^2$ level of the laser intensity. The averaged power of the pump laser was 7~\si{\milli\watt} and that of the probe laser 13~\si{\milli\watt}. For the imaging experiments, the sample was mounted on a X-Y positioning stage equipped with step motors ensuring the positioning precision of 0.16~\si{\micro\metre}. The used step, for the here-presented images, was 1.25~\si{\micro\metre}.

\subsection*{Processing of the transient reflectivity signals to get the TDBS signals}
The TDBS signals shown in Fig.~\ref{fig:typical_signals} are obtained from the raw transient reflectivity signals captured by the photodetector followed by filtering, subtracting the thermal background and cutting of initial peaks generally observed at the time scale of the pump and probe laser pulses overlapping. First the signal is filtered with a 48th-order FIR bandpass filter with passband $10<f<50$~\si{\giga\hertz} according to the expected Brillouin frequencies from Tab.~\ref{tab:properties}. Then, the remaining after filtering background was subtracted applying the local regression method LOESS from Matlab\textsuperscript{\textregistered} using weighted linear least squares and a 2nd degree polynomial model\cite{MatlabDoc} with 319 points of span of the moving average (\textit{i.e.}, the window span for the moving average is of about 0.15~\si{\nano\second}). As the fast non-oscillating transients of optical reflectivity near the initial time are difficult to remove without influencing on the high-frequency Brillouin oscillations (Brillouin peak), the first 25~\si{\pico\second} of the remaining signal were cut. Therefore, the first tens of nanometres next to the iron generator are lost for the analyses. We applied a similar treatment to all signals presented in this paper before further processing.

\subsection*{Estimation of the expected Brillouin frequencies in both phases VI and VII of \chemform{H_2O} ice at 2.15~\si{\giga\pascal}}
At the room temperature and the pressure of 2.15~\si{\giga\pascal}, both phases VI and VII of \chemform{H_2O} ice are coexisting (phase transition)\cite{Polian1983_v27_p6409--6412, Shimizu1996_v53_p6107--6110, Shimizu1996_v219-220_p559---561, Baer1998_v108_p4540-4544, Dunaeva2010_v44_p202--222}. We give here details on the way the estimates of the Brillouin frequencies have been done for both phases. The goal was to have at hand estimated values of the expected Brillouin frequencies for the subsequent analysis of the experimental data. The error of the estimates is overall not very small, especially because of the drastic change of properties the sample material can exhibit at the phase transition. We do believe though that the below presented strategy of extrapolation allows anyway the 3D characterisation of individual ice grains at high-pressure and in two-phase region with an acceptable correctness.

The density, reported in the first row of Tab.~\ref{tab:properties}, are deduced by interpolating the same data than the ones used in Fig. 3 of Ref.~\citeonline{Polian1983_v27_p6409--6412} plotting the density of the various \chemform{H_2O} forms as a function of pressure. The uncertainties in the values of the densities in Ref.~\citeonline{Polian1983_v27_p6409--6412} and also used in Ref.~\citeonline{Shimizu1996_v53_p6107--6110}, wherefrom the elastic constants can be taken, are negligible in comparison with the uncertainties of the elastic moduli. The elastic constants are deduced by fitting and extrapolating to a pressure of 2.15 GPa the results presented in Figs. 4 and 5 of Ref.~\citeonline{Shimizu1996_v53_p6107--6110}. Thus, the accuracy of $\sim\pm2$\%, reported in Ref.~\citeonline{Shimizu1996_v53_p6107--6110} for the ratios of the elastic moduli and the density, results in $\sim\pm2$\% accuracy in the moduli and in $\sim\pm1$\% accuracy in the velocities of the acoustic modes\cite{Shimizu1996_v53_p6107--6110}. The refractive index of ice was evaluated in Ref.~\citeonline{Shimizu1996_v53_p6107--6110} (Fig. 3) with the accuracy of $\sim\pm2$\%.

To calculate the Brillouin frequency ranges, the density and the elastic constants are used to solve the Christoffel equation in all possible propagation direction to get the minimum and maximum possible values of velocity for each acoustic mode in each phase of \chemform{H_2O} ice. Then Eq.~\eqref{eq:BrillouinFrequency} is used to get the minimum and maximum possible values of Brillouin frequency in each case. Since the acoustic velocities are known in any propagation direction of the CAPs, so do the Brillouin frequencies. Combining the uncertainties in velocities and in the refractive indices, we could conclude that maximum and minimum Brillouin frequencies in Table~\ref{tab:properties} and Fig. S1 (see Supplementary information) are determined with the uncertainties of $\sim\pm3$\%. The uncertainty of the Brillouin frequencies confining the overlap interval of the LA Brillouin frequencies in ices VI and VII, i.e. $[27.5, 29.4]$~\si{\giga\hertz}, can be scaled respectively. Thus, the overlap interval shrinks but does not disappear ($[27.5, 29.4]$~\si{\giga\hertz} $\Rightarrow$ $[28.3, 28.5]$~\si{\giga\hertz}).

%\bibliography{Bib_SciRep2020}

\section*{Acknowledgements}

This work was supported by the French National Research Agency (ANR, France) through the grant <ANR-18-CE42-017>. T.T. was supported by the Région Pays de la Loire through the RFI Le Mans Acoustique (project \enquote{Paris Scientifique OPACOP 2018}). E.D.L.S. was supported by the program Acoustic Hub\textsuperscript{\textregistered} funded by the Région Pays de la Loire. We thank our colleagues from NETA who provided insight and expertise that greatly assisted the research.%The authors thank Dr Maju Kuriakose for collaboration and discussions.

\section*{Author contributions statement}

A.Z.,V.E.G., S.R., N.C. and A.B. designed the research, N.C. and A.Z. prepared the sample, S.S., E.D.L.S, S.R., N.C. and A.B contributed to the experiments, S.R., T.T., S.S., E.D.L.S and V.T. contributed to signal processing, V.E.G., S.R., T.T. ans S.S. developed the theoretical estimates, S.R., V.E.G., A.Z., S.S., N.C., A.B. and V.T. analysed and interpreted the experimental observations, S.R., V.E.G. and A.Z. wrote the manuscript. All authors reviewed the manuscript.

\section*{Additional information}

The authors declare no competing interests.

\newpage

\textbf{\Huge Supplementary information}

\section*{Supplementary note on the distribution of the Brillouin frequency for phases VI and VII of \chemform{H_2O} ice}
\begin{figure}[htb!]
\centering
\includegraphics[width=\linewidth]{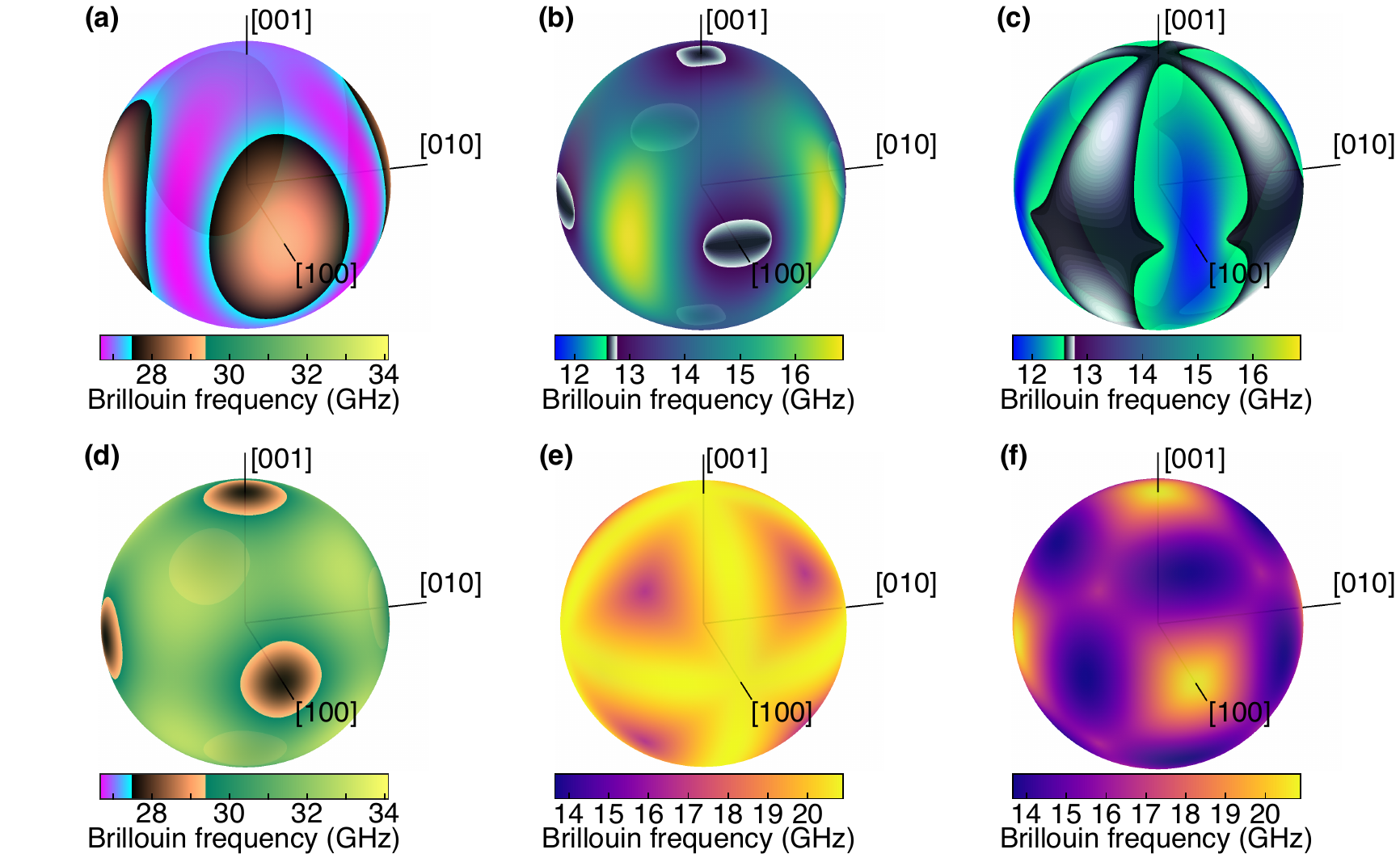}%[width=85mm]
\caption{Distribution of the Brillouin frequency of each acoustic mode for (top) the phase VI and (bottom) the phase VII of \chemform{H_2O} ice along the principal axes of the crystals as a function of the orientation of the normal to the iron/ice interface relative to the ice crystal principal axes. The results are presented separately for different modes: in (a,d) for the LA mode, in (b,e) for the FTA mode, and in (c,f) for the STA mode.}
\label{fig:S1}
\end{figure}

The Brillouin frequency (BF) ranges for both phases are calculated as follows. The Christoffel equation is solved (see Section 4.2 in Ref.~\citeonline{Royer1999_v1_p}), using the density and the elastic constants reported in Tab. 1 of the manuscript, in all possible propagation direction to get the minimum and maximum possible values of velocity for each acoustic mode in each phase of \chemform{H_2O} ice. Reminding the relation between the BFs, $f_{B,\alpha}$, and the acoustic velocities, $\varv_\alpha$ (with $\alpha$ the type of the acoustic mode: LA or one of the TA) [Eq. (1) of the manuscript]:
\begin{align}\label{eq:BrillouinFrequencySupp}
    f_{B,\alpha} = \frac{2n\varv_\alpha}{\lambda_\text{probe}}\,,
\end{align}
 where $n$ is the refractive index of the transparent media at the wavelength in vacuum $\lambda_\text{probe}$ of the probe laser pulses, the minimum and maximum possible values of BF can be calculated for each mode of both phases. Note that the optical anisotropy of the phase VI that is reported to be less than 1\%\cite{Polian1983_v27_p6409--6412} is neglected. Since the acoustic velocities are known in any propagation direction of the CAPs, so do the BFs, which allows to plot the surfaces shown in Fig.~\ref{fig:S1} that depict the theoretical distribution of the BF of each acoustic mode as a function of the propagation direction of the CAP for (top) the phase VI and (bottom) the phase VII of \chemform{H_2O} ice. The results are presented separately for different modes: in (a,d) for the LA mode, in (b,e) for the FTA mode, and in (c,f) for the STA mode.
 %%%%%%%%%%%%%%%%%%%%%%%%%%%%%%%%%%%%%%%%%%%%%%%%%
%%%%%%%%%%%%%%%%%%%%%%%%%%%%%%%%%%%%%%%%%%%%%%%%%

 The different symmetries of both phases of water ice (tetragonal for phase VI and cubic for phase VII) are visible in Fig.~\ref{fig:S1} in the symmetries of the color distribution on the surfaces of the spheres. The color scale of Fig.~\ref{fig:S1}(a,d) is the same and has been chosen to be the same as in the Figs.~3(a) and 4 of the manuscript showing the 2D and 3D images with the LA mode, respectively. This color scale has been designed to quickly depict whether the value of the BF of the LA mode is lower than 27.5~\si{\giga\hertz} (magenta-to-cyan color scale), in which case it could only be attributed to phase VI, or higher than 29.4~\si{\giga\hertz} (greenish color scale), in which case it could only be attributed to phase VII. Where the frequency ranges overlap between both phases ($f_{B,\text{LA}}\in[27.5,29.4]$~\si{\giga\hertz}), the color scale is copperish. This designed scale allows in a glance to inform about the directions in both phases that share the same LA Brillouin frequency and therefore in which case additional information from TA modes (presence/absence in the signal, value of the BF) is mandatory to recognise the phase (see Discussion in the manuscript). The color scale of Fig.~\ref{fig:S1}(b,c) is the same and has also been designed to show the directions of propagation in which the BF could be either that of the FTA mode or that of the STA mode. Note that, obviously, the STA velocity remains always smaller than the FTA one. The color scale of Fig.~\ref{fig:S1}(e,f) is the same and has been chosen to be the same as in the Fig.~3(b) of the manuscript. Since most of the detected TA mode are the slow one in phase VII (see Discussion in the manuscript), this choice allows in a glance to qualitatively inform about the orientation, relatively to the diamond anvil cell axis, of the probed grains exhibiting TA modes of particular frequency.

\section*{Supplementary note on the coherence length of the probe laser pulses}
The preliminary estimates, described in the main text, demonstrate that the depth of the TDBS imaging in our experiments could be limited rather by the coherence length of the probe laser pulses than the other factors such as the probe laser beam diffraction length or the acoustical phonons and probe light photons penetration depths.

\begin{figure}[htb!]
\centering
\includegraphics[width=\linewidth]{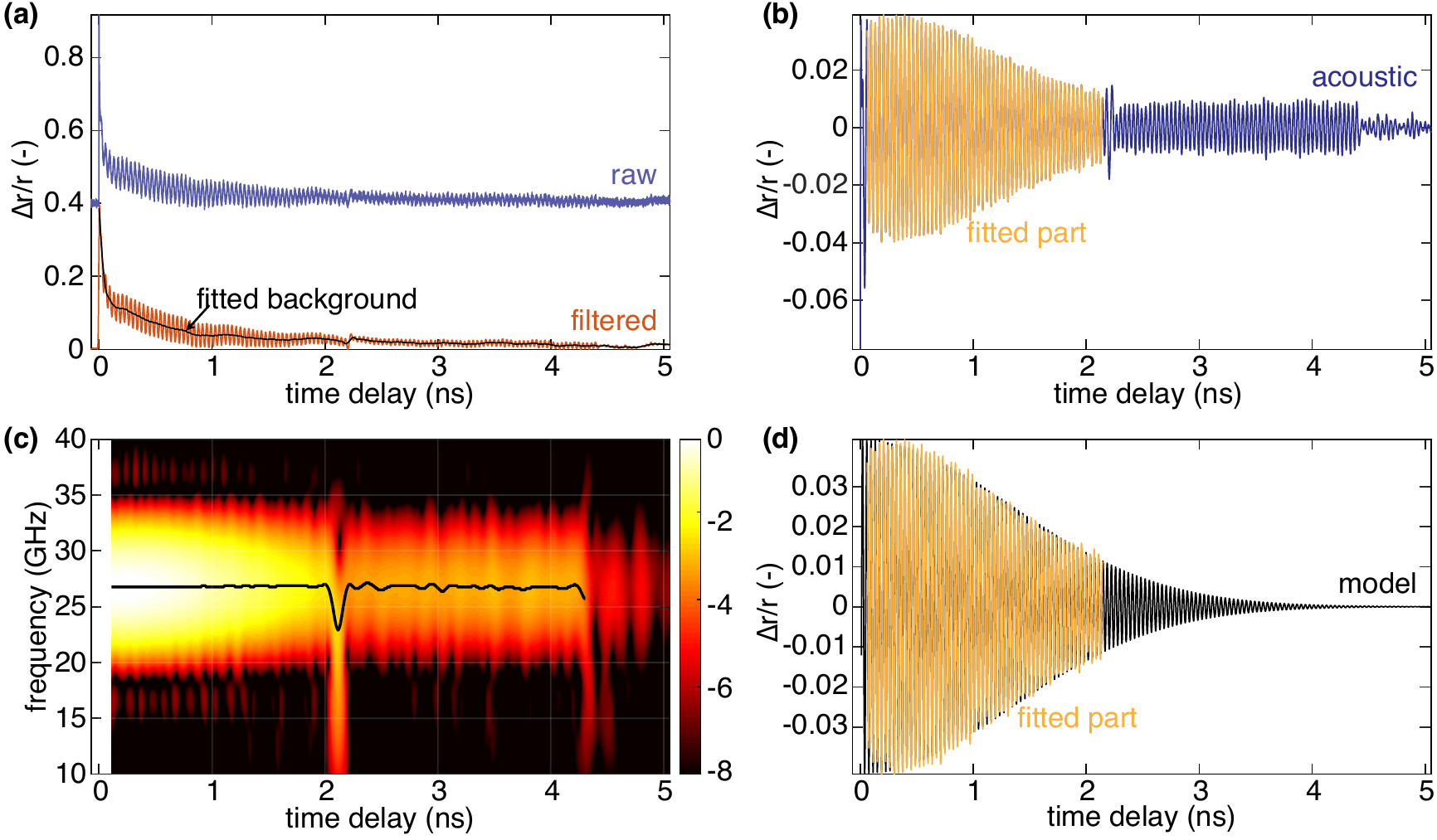}%[width=85mm]
\caption{(a) Transient reflectivity signal at a position around $x= 47.5$~\si{\micro\metre} and $y=85$~\si{\micro\metre} as a function of time delay: raw data (top), filtered data (bottom) and fitted background (black line); the raw data are artificially upshifted for clarity purpose. (b) Acoustic contribution to the transient reflectivity signal shown in (a) as a function of time delay; this signal corresponds to the subtraction of the fitted background from the filtered signal. (c) Time-frequency representation (TFR) of the acoustic contribution shown in (b). (d) Comparison of the fitted part of the acoustic contribution in (b) with the model signal resulting from the fitting.}
\label{fig:S2}
\end{figure}

Figure~\ref{fig:S2}(a) presents the transient reflectivity signal, as a function of time delay, of the longest duration revealed in our TDBS imaging experiments at a position around $x= 47.5$~\si{\micro\metre} and $y=85$~\si{\micro\metre}. The raw data (top) as well as the filtered data (bottom) and the fitted background (black line) are shown. Note that the processing of the raw transient reflectivity signal is exactly the same than the one described and used within the manuscript. The estimates provide opportunity to relate this signal to the propagation of the CAP from the iron optoacoustic transducer to the interface of the ice and the diamond anvil through a single crystallite of ice [see the schematic presentation of the sample in Fig. 1(c)]. The propagation time of about 2.15~\si{\nano\second} until the CAP reflection at the ice/diamond interface is revealed from the structure of the processed signals in Fig.~\ref{fig:S2}(a)-(c). Figure~\ref{fig:S2}(b) presents the acoustic contribution to the transient reflectivity signal obtained by subtracting the thermal background and other non-oscillating contributions [fitted background in Fig.~\ref{fig:S2}(a)] from the raw signal. Time-frequency representation (TFR) of the acoustic contribution [Fig.~\ref{fig:S2}(b)] is estimated by a short-time Fourier transform (STFT) of the signal with a 0.23~\si{\nano\second} Hann window and is shown in Fig.~\ref{fig:S2}(c). The scale of the spectrogram and associated color bar is logarithmic for reading purpose. A ridge extraction is performed by estimating the frequency of the peak in the Fourier domain for each temporal position of the Hann window of the STFT [black solid line in Fig.~\ref{fig:S2}(c)]. The frequency down and up shift at $t\approx 2.15$~\si{\nano\second} on that ridge is attributed to the reflection of the CAP at the ice/diamond interface: the sudden change of phase in the oscillating signal, \textit{i.e.} the discontinuity due to the reflection of the coherent acoustic pulse, indeed leads to an instant in the TFR when the frequency content span over a wider range.
%%%%%%%%%%%%%%%%%%%%%%%%%%%%%%%%%%%%%%%%%%%%%%%%%
%%%%%%%%%%%%%%%%%%%%%%%%%%%%%%%%%%%%%%%%%%%%%%%%%

Figure~\ref{fig:S2}(d) illustrates the fitting of a selected part of the acoustic signal with a sinusoidal gaussian-decaying model signal: $A\exp(-\alpha t)\exp[-2(t/\tau)^2]\cos(2\pi f_{B,\text{LA}} t + \phi)$, with $A$ the amplitude, $\alpha$ the acoustic absorption coefficient, $\tau$ the coherence time of the probe laser pulses, $f_{B,\text{LA}}$ the BF of the detected LA mode, and $\phi$ the phase of the signal. Note that $A$ and $f_{B,\text{LA}}$ are not fitted but considered as fixed constants. $A$ is equal to the maximum amplitude of the fitted part. $f_{B,\text{LA}}$ is deduced from the Fourier spectrum density of the selected part of the signal depicted by the yellow line in Fig.~\ref{fig:S2}(b): $f_{B,\text{LA}}=26.85$~\si{\giga\hertz} [see Fig.~\ref{fig:S2}(c)]. The selected part of the acoustic signal avoids purposely the reflection of the CAP at $t\approx 2.15$~\si{\nano\second}. The estimated coherence time from the fitting is $\tau=2.61$~\si{\nano\second} with 95\% confidence bounds of 2.59 and 2.62~\si{\nano\second}. Multiplication of $\tau$ by the the LA mode velocity ($\varv_\text{LA}=\frac{f_{B,\text{LA}}\lambda_\text{probe}}{2n}$ , see Eq.~\ref{eq:BrillouinFrequencySupp}) leads to the estimate of the coherence length of $L_\text{coherence}^\text{probe}=12.7$~\si{\micro\metre} with 95\% confidence bounds of 12.64 and 12.79~\si{\micro\metre}. The relation between the coherence time $\tau$ and the duration of the probe laser pulses $\tau_\text{probe}$ reads
\begin{align}
    \tau=\frac{1}{2}\left(\frac{c_0/n}{\varv_\text{LA}}\right)\tau_\text{probe}=\frac{c_0\tau_\text{probe}}{f_{B,\text{LA}}\lambda_\text{probe}}\,,
\end{align}
where $c_0$ denotes the speed of light in vacuum. The duration of the probe laser pulses is therefore estimated to be equal to $\tau_\text{probe}=124$~\si{\femto\second}. For the estimates here, we have used the refractive index of the phase VI of ice water $n=1.468$ and the sound velocity $\varv_\text{LA}=4881$~\si{\metre\per\second} derived from the measured BF. Both the coherence length and the duration of the laser pulse are close to the values obtained by the simplest estimates in the main text. The obtained fit is not improved by including the acoustic absorption since the fitted value for $\alpha$ is 0, or more precisely is equal to the limit bound (0) within numerical error range. Moreover, the fit cannot be improved/modified by including either corrections due to the light diffraction or to the optical absorption. This demonstrates the dominance of the coherence effect influence on the dynamics of the Brillouin oscillation amplitude in our experiments.

\section*{Supplementary note on the micro-Raman spectrometry of the water ice sample}
Micro-Raman spectrometry allows performing studies of materials with micrometric spatial resolution, even from a distance, and across any optically transparent window up to several millimetres in thickness. In these conventional Raman conditions (\textit{i.e.}, apart from near-field Raman techniques), the spatial resolution is driven by confocal systems. In the best conditions, the lateral resolution is about 0.4~\si{\micro\metre}, and the axial resolution is a few micrometres, and it becomes larger when the window thickness increases, as it occurs in a diamond anvil cell (DAC)\cite{Mao1986_v91_p4673-4676,Everall2010_v135_p2512-2522,Bruneel2002_v33_p815-828,Graf2007_v7_p238-242}.

\begin{figure}[htb!]
\centering
\includegraphics[width=\linewidth]{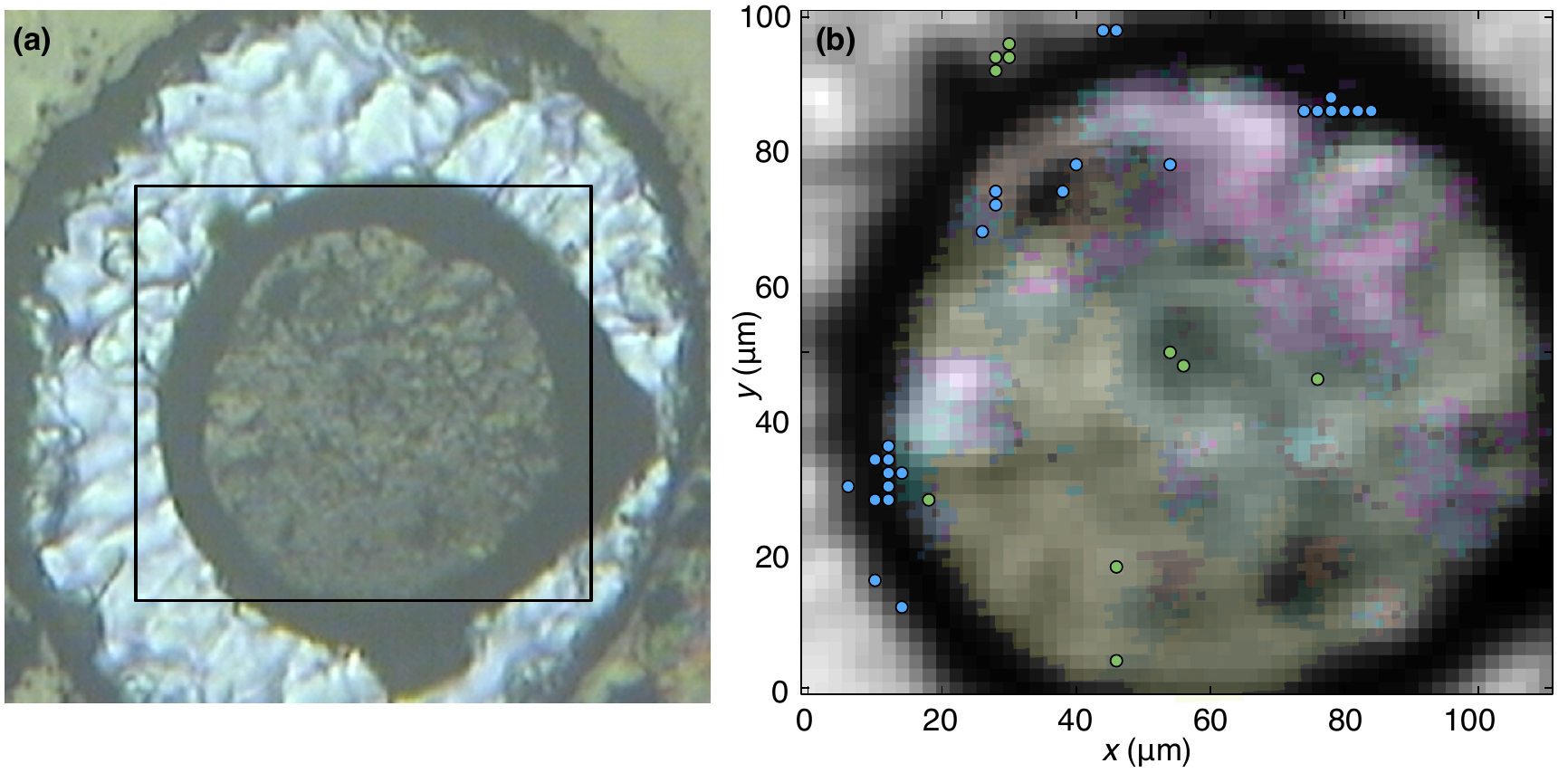}%[width=85mm]
\caption{(a) Image of the sample (captured using $\times25$ magnification objective) showing the iron generator surrounded by transparent ice. The black rectangle stands for the area scanned with the spectrometer. (b) Optical mapping of the sample under white illumination in gray color scale. The blue dots show all the areas where the phase VI of \chemform{H_2O} ice alone is observed ; the green dots show selected areas where almost pure phase VII is observed. The color tones have been obtained by superposing to the gray color scale the 2D image of the Fig.~3(a) of the manuscript in transparency.}
\label{fig:S3}
\end{figure}

\begin{figure}[htb!]
\centering
\includegraphics[width=\linewidth]{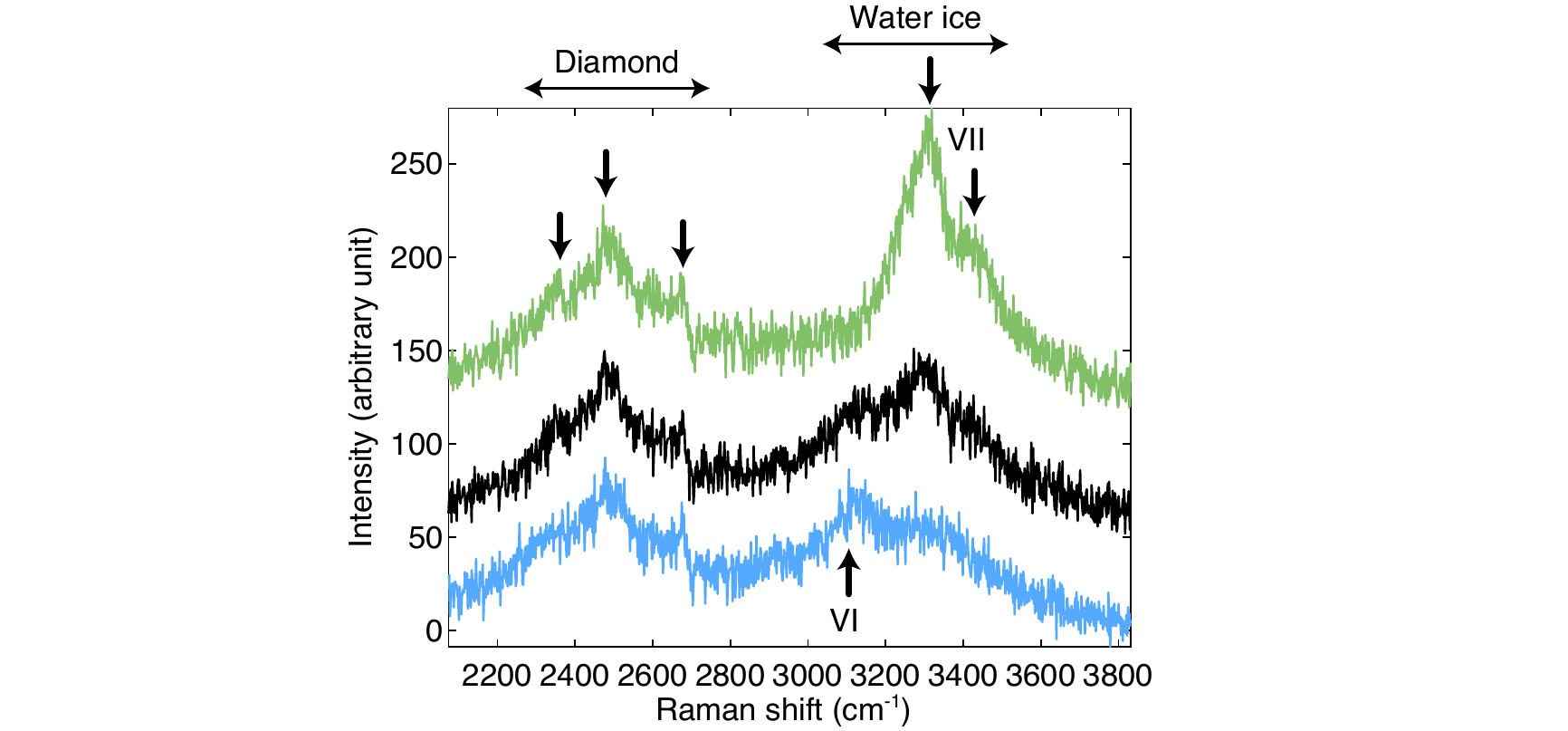}%[width=85mm]
\caption{Raman spectra collected in the DAC, characteristic of phase VI (bottom blue), phase VII (top green) or a mixture of the two phases (middle black).}
\label{fig:S4}
\end{figure}

Raman spectra were measured here to identify the ice phases present in the sample\cite{Hsieh2015_v5_p8532}. A $100\times110$~\si{\micro\metre\squared} mapping with a 2~\si{\micro\metre} step has been done to get the spatial distribution of each phase around and over the generator. The spectra were collected with a T64000 Horiba-Jobin-Yvon spectrometer, under microscope with $\times25$ objective, using the 514.5~\si{\nano\metre} wavelength radiation of an \chemform{Ar}-\chemform{Kr} laser as excitation, and less than 5~\si{\milli\watt} power on the sample.  The sample and its optical mapping are shown in Fig.~\ref{fig:S3}, and selected Raman spectra are plotted in Fig.~\ref{fig:S4}. The image of the sample shown in Fig.~\ref{fig:S3}(a) was captured using $\times25$ magnification objective and depicts the iron generator surrounded by transparent ice. The black rectangle stands for the area scanned with the spectrometer. In Fig.~\ref{fig:S3}(b), the optical mapping of the sample collected with the spectrometer under white illumination is shown in gray color scale: black traduces a low intensity, white a high one. The intensity for each pixel of the map is calculated, \textit{i.e.}, integrated, over the whole spectral range collected by the spectrometer photodetector.
%%%%%%%%%%%%%%%%%%%%%%%%%%%%%%%%%%%%%%%%%%%%%%%%%
%%%%%%%%%%%%%%%%%%%%%%%%%%%%%%%%%%%%%%%%%%%%%%%%%

%%%%%%%%%%%%%%%%%%%%%%%%%%%%%%%%%%%%%%%%%%%%%%%%%
%%%%%%%%%%%%%%%%%%%%%%%%%%%%%%%%%%%%%%%%%%%%%%%%%

In Fig.~\ref{fig:S4}, the spectra are characteristic of a mixture of phase VI and phase VII of \chemform{H_2O} ice, as reported in Ref.~\citeonline{Hsieh2015_v5_p8532}: main peaks expected at about $3130$~\si{\per\centi\metre} Raman shift for the phase VI and at about $3300$~\si{\per\centi\metre} and $3400$~\si{\per\centi\metre} Raman shifts for the phase VII. In addition, several areas show extra high intensity signals attributed to luminescence of ruby dust (introduced for pressure measurement in the DAC) and presumably (no unambiguous evidence found) of iron ions \chemform{Fe^{2+}} and \chemform{Fe^{3+}} that are often generated in presence of water with dissolved oxygen, especially after heating as induced by lasers. These high intensity signals of luminescence hide the weak Raman signals in many places preventing a full distribution mapping of the two phases of water ice over the surface of the generator using Raman spectrometry. We report instead, in Fig.~\ref{fig:S3}(b), areas where the phase VI alone exists marked with the blue dots. Outside these positions, the Raman signals, where treatable, are characteristic of phase VII (green dots) or of a mixture of both phases VI and VII of water ice. These distributions are in very good agreement with the TDBS analysis reported in Fig.~3(a) of the manuscript, although the TDBS technique provides, in addition, the distribution along the depth of the phases with a better axial resolution (shown in Fig.~4 of the manuscript). The color tones of Fig.~\ref{fig:S3}(b), obtained by superposing to the gray color scale of the optical mapping the 2D image of the Fig.~3(a) of the manuscript in transparency, is provided for ease of comparison. Please note that the origins of Fig.~3(a) and Fig.~\ref{fig:S3}(b) were different, that the required translation was done by manual adjustment and that, by comparing the optical images from both devices, a slight angular rotation of \ang{2.05} clockwise was performed.

\end{document}